\documentclass[useAMS,usenatbib]{mn2e}
\usepackage{amsmath}
\usepackage[dvips]{graphicx}
\usepackage{url}
\title[Type Ia Supernova SN 2012dn]{Supernova SN 2012dn: A spectroscopic clone of SN 2006gz}

\author[N. K. Chakradhari et al.]
{N. K. Chakradhari$^1$, D. K. Sahu$^2$,
S. Srivastav$^2$, G. C. Anupama$^2$\\
1. School of Studies in Physics \& Astrophysics, Pt. Ravishankar Shukla University, Raipur 492010, India\\
2. Indian Institute of Astrophysics, Koramangala, Bangalore 560 034, India\\
E-mail : nkchakradhari@gmail.com (NKC), dks@iiap.res.in (DKS), gca@iiap.res.in (GCA)}

\begin{document}

\date{Accepted .....; Received ......}

\pagerange{\pageref{firstpage}--\pageref{lastpage}} \pubyear{2014}

\maketitle

\label{firstpage}
\begin{abstract}
We present optical and UV analysis of the luminous type Ia supernova SN 2012dn 
covering the period $\sim -$11 to +109 days with respect to the $B$ band maximum, 
that occurred on 
JD 2456132.89 $\pm$ 0.19, with an apparent magnitude of $m_{B}^\text{max}$ = 14.38 
$\pm$ 0.02.  The absolute magnitudes at maximum in $B$ and $V$ bands are 
$M_{B}^\text{max} = -19.52 \pm 0.15$  and $M_{V}^\text{max} = -19.42 \pm 0.15$, 
respectively. SN 2012dn is marginally luminous compared to normal type Ia supernovae.
The peak bolometric luminosity of $\log L_\text{bol}^\text{max} = 43.27 
\pm 0.06$ erg s$^{-1}$ suggests that $0.82 \pm 0.12$ M$_\odot$ of $^{56}$Ni 
was synthesized in the explosion. The decline rate $\Delta m_{15}(B)_\text{true}= 0.92 \pm 0.04$ 
mag is lower than that of normal type Ia supernovae, and similar to 
the luminous SN 1991T. However, the photometric and spectroscopic behaviour of 
SN 2012dn is different from that of SN 1991T. Early phase light curves in $R$ 
and $I$ bands are very broad. The $I$ band peak has a plateau-like appearance 
similar to the super-Chandra SN 2009dc. Pre-maximum spectra show clear evidence 
of C\,{\sc ii} 6580 \AA\, line, indicating the presence of unburned materials. The velocity evolution of C\,{\sc ii} line is peculiar.  Except for the very early phase ($\sim-$13 d),
the  C\,{\sc ii} line velocity is lower than the velocity estimated using the Si\,{\sc ii} line.
During the pre-maximum and close to maximum phase, to reproduce observed shape of 
the  spectra, the synthetic spectrum code {\sc syn++}  
needs significantly higher blackbody temperature than those required for normal type Ia events.
 The photospheric velocity evolution and other spectral properties are similar to 
those of the carbon-rich SN 2006gz.
\end{abstract}


\section{Introduction}
\label{sec:intro}
Type Ia supernovae (SNe Ia), characterized by the presence of the Si\,{\sc ii} 
6355 \AA\, absorption feature, are believed to be thermonuclear explosions of 
carbon/oxygen (C/O) white dwarf (WD) either accreting matter from its companion
in a binary system, or merging with another white dwarf \citep{hil00,lei00,
fil97}. A majority of these objects, known as `normal' type Ia SNe, 
\citep{bra06} form a remarkably homogeneous class with similar light curve 
shape and spectral evolution. The high and uniform peak luminosity makes them 
powerful distance indicators. SNe Ia have been used to measure the cosmic 
expansion history, leading to the discovery of accelerating universe and 
existence of dark energy \citep{rie98,per99}. The key fact behind this is that 
the light curves of normal SNe Ia can be normalized to a common light curve 
using empirical relations between light curve width and peak luminosity 
\citep*{phi93,rie96,per97,phi99,gol01}.

The observed homogeneity in SNe Ia is linked with the commonly accepted theory 
that the explosion of WD occurs at a mass close to the Chandrasekhar mass limit 
\citep{cha31} of 1.4 M$_\odot$. However, the detailed explosion mechanism and 
the nature of progenitors are still under investigation. With the availability 
of good quality data of  SNe Ia in the recent years, deviation from the observed
homogeneity, and the occurrence of peculiar and exceptional events is quite 
evident. A recent study by \citet{liw11} has revealed that a majority of SNe Ia
($\sim$ 70\%) belong to a fairly homogeneous class -- `normal' type Ia, while 
$\sim$ 9\% are overluminous -- SN 1991T-like events \citep{fil92a,phi92}, 
$\sim$ 15\% are underluminous -- 1991bg-like events \citep{fil92b,lei93} and 
$\sim$ 5\% are  peculiar -- SN 2002cx-like events \citep{jha06,sah08,fol13}. 

The discovery of the extremely luminous type Ia supernova SN 2003fg 
\citep[SNLS-03D3bb:][]{how06} and the subsequent discoveries of SN 2006gz, 
SN 2007if and SN 2009dc challenge the existing theory of type Ia SN explosion 
of Chandrasekhar mass WD \citep[and references therein]{hic07,yam09,sca10,
sil11,tau11}. Assuming the luminosity of SNe Ia is powered by radioactive decay 
of $^{56}$Ni synthesized in the explosion, the mass of $^{56}$Ni required to 
produce the observed luminosity of this class of objects exceeds 1 M$_\odot$. 
Within the spherically symmetric explosion scenario, the existing models of SN Ia explosion of Chandrasekhar mass WD cannot produce the required 
$^{56}$Ni. The inferred mass of $^{56}$Ni can be produced only with a progenitor
more massive than a Chandrasekhar mass WD. Hence, these extremely luminous type 
Ia SNe have been termed as `super-Chandrasekhar-mass', or `super-Chandra' type Ia.
They have very bright peak at maximum $M_{V}^\text{max}$ $\sim -$20 mag and the 
early post-maximum decline is very slow. Apart from moderately low ejecta 
velocity and presence of unburned carbon, the near-maximum spectra of these 
objects are similar to those of normal SNe Ia. At nebular phase these SNe 
exhibit a photometric and spectroscopic behaviour diverse from normal SNe Ia.
They show  enhanced  fading in the light-curve with a possible 
 break at  $\sim$ 150--200 d, and very weak/absence of 
[Fe\,{\sc iii}] emission in their nebular spectra \citep{mae09,tau13}. 
The luminous and broad light curves of the  super-Chandra SNe do not 
follow the  Phillips luminosity-decline relation \citep{phi93,phi99} 
and  may  lead to wrong conclusions if included in cosmological 
studies \citep{tau11}. 

To explain the observed properties of different classes of SNe Ia, a variety of 
theoretical models, incorporating both single-degenerate (SD), and 
double-degenerate (DD) progenitors, with their masses ranging from 
sub-Chandrasekhar to super-Chandrasekhar, and propagation of the thermonuclear 
burning front via deflagration (subsonic) to detonation (supersonic) have been 
proposed \citep[and references therein]{how11}. Detonation  models produce 
mostly Iron Group Elements (IGEs), while deflagration models produce both Intermediate
Mass Elements (IMEs) and IGEs, consistent with the observed  spectra of SNe Ia. 
Deflagration models leave most of the materials unburned, and are unable to 
produce high velocity features. The delayed detonation model \citep*{kho91,gam05}, 
in which a deflagration to detonation transition occurs at some stage of the 
thermonuclear explosion, is however, successful in reproducing the observed 
characteristics of SNe Ia. Models incorporating broken symmetries \citep*{kas09}
and asymmetric explosion \citep{mae09I,mae10} have also been proposed to account 
for the observed diversity among SNe Ia. 

In a sample of 29  objects, \citet{ham96} found that most of
the luminous SNe Ia were found in galaxies with young stellar environments. With a 
larger sample of SNe Ia, \citet{sul10} have shown that their light curve width
closely tracks the specific star formation rate (sSFR) and stellar mass of the 
host galaxies -- massive and/or low sSFR galaxies host SNe with lower stretches 
(narrow light curves or higher $\Delta m_{15}(B)$). Super-Chandra SNe Ia are 
also suggested to occur preferentially in young stellar and low metallicity 
environments \citep{tau11,kha11}. Differential rotation and merging mechanisms 
may play key roles in increasing the progenitor mass suitable for these objects.
A rotating WD can support more mass than the classical Chandrasekhar mass limit 
of non rotating WD, with differentially rotating WDs being more massive than the
rigidly rotating WDs \citep{yoo05}.

In this paper, we present optical and UV data
of the luminous type Ia supernova SN 2012dn observed with 2-m Himalayan Chandra
Telescope (HCT), Hanle and {\it Swift} UVOT. 
Section \ref{sec:observation} 
describes the observation and data reduction techniques. Photometric and 
spectroscopic results are presented in Section \ref{sec:light_curve_analysis} and \ref{sec:spec_evol}. We discuss and summarize the paper in Section \ref{sec:summary}.

\begin{figure}
\begin{center}
\resizebox{0.80\hsize}{!}{\includegraphics{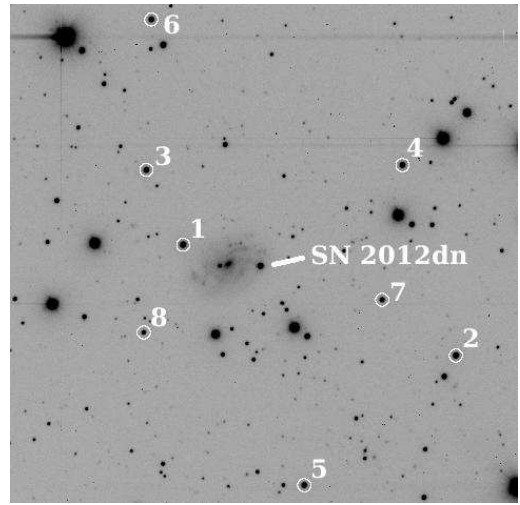}}
\caption[]{$R$ band image of SN 2012dn in the galaxy ESO 462-016. The field 
stars marked with numbers 1--8 are used as local standards, and their calibrated
magnitudes are listed in Table \ref{tab_std}. North is up and east to the left.
The field of view is 10 $\times$ 10 arcmin$^2$.}
\label{fig_std}
\end{center}
\end{figure}

\section{Observation and Data reduction}
\label{sec:observation}
\begin{table}
\caption{Properties of SN 2012dn and its host galaxy.}
\begin{tabular}{llc}
\hline\hline     
\textbf{SN 2012dn} && Ref. \\
\\
$\alpha$ (J2000) & 20$^\text h$23$^\text m$36$^\text s$.26& 1\\ 
$\delta$ (J2000) &$-$28$^\circ$16$'$43$''$.40& 1\\
SN offset from galaxy nucleus & 35$''$ W, 3$''$ S & 1\\
SN type & Ia & 1 \\
$E(B-V)_\text{host}$ & 0.12 & 2 \\
$E(B-V)_\text{Gal}$ & 0.06 & 3 \\
Discovery date (UT) & 2012 July 08.52 & 1\\
Discovery JD & 2456116.52 & 1\\
Discovery magnitude & 16.3 & 1\\
JD$_B^\text{max}$ & 2456132.89 & 2\\
$\Delta m_{15}(B)_\text{true}$ & 0.92 &2\\
$m_{B}^\text{max}$ & 14.38 &2\\
$M_{B}^\text{max}$ ($\mu$) & $-$19.52 & 2\\
Mass of $^{56}$Ni & 0.82 M$_\odot$ & 2\\
\\
\textbf{ESO 462-16} & &\\
\\
PGC name & PGC 64605&3\\
Galaxy type & SA(s)cd &3\\
$\alpha$ (J2000) & 20$^\text h$23$^\text m$38$^\text s$.9& 3\\ 
$\delta$ (J2000) &$-$28$^\circ$16$'$40$''$&3\\
Redshift (\textit{z}) & 0.010 &3 \\
Extn. Corr. app. B mag. & 13.05 &4\\
Absolute B mag ($\mu$) & $-$20.10 & 4,5 \\
Diameter& 1.7$'$ $\times$ 1.2$'$& 3\\
$v_\text{Virgo}$&3070 km\,s$^{-1}$&3\\
$\mu$& 33.15 &5\\
$E(B-V)_\text{Gal}$ & 0.06 & 3\\
\hline
\multicolumn{3}{p{7.4cm}}{ Ref.: 1. \citet{boc12}; 2. this paper \newline 
3. NED (\url{http://nedwww.ipac.caltech.edu}) \newline 
4. LEDA (\url{http://leda.univ-lyon1.fr}) \newline
5. using $H_0$ = 72 km\,s$^{-1}$\,Mpc$^{-1}$ on $v_\text{Virgo}$}
\end{tabular}
\label{tab_host}

\end{table}

SN 2012dn (PSN J20233626-2816434) was discovered by Stuart Parker on 2012 July 
08.52 (UT) at a red magnitude of 16.3 in the SA(s)cd type galaxy ESO 462-G16 
(PGC 64605), at a redshift of $z$ = 0.010 \citep[source NED]{the98}. The 
supernova was located (Fig. \ref{fig_std}) at $\alpha$ = 20$^\text h$23$^\text m$36$^\text s$.26 (J2000) and $\delta$ = $-$28$^\circ$16$'$43$''$.40 (J2000), 
 35$''$ west and 3$''$  south of the nucleus of the host galaxy \citep*{boc12}. 
Based on a spectrum obtained on July 10.2 (UT) with the 8-m Gemini South 
Telescope, \citet{par12} classified it as a type Ia supernova, approximately a 
week before maximum light. The blueshift of the Si\,{\sc ii} absorption minimum 
was measured as 12000 km\,s$^{-1}$. A strong C\,{\sc ii} 6578 \AA\, absorption 
feature, blueshifted  by $\sim$ 12200 km\,s$^{-1}$ was detected in the spectrum.
Based on the similarity of the spectrum of SN 2012dn with the pre-maximum 
spectra of SN 2006gz \citep{hic07}, SN 2007if \citep{sca10} and SN 2009dc 
\citep{yam09}, it was suggested to be a possible super-Chandra event \citep{par12}. 
Analyzing the spectrum obtained on July 11.5 (UT) with the 2-m telescope of 
University of Hawaii, \citet{cop12}  confirmed it to be a peculiar type Ia event, similar to    
SN 2009dc around 10 d before peak brightness. The velocity and spectral slopes 
were found to be very similar to SN 2006gz at a similar epoch. Some basic 
information of SN 2012dn and its host galaxy ESO 462-G16 are listed in Table 
\ref{tab_host}.

\subsection{Photometry}
\subsubsection{Optical Photometry using 2-m HCT}
Observations of SN 2012dn were carried out using the Himalaya Faint Object 
Spectrograph Camera (HFOSC) mounted on the 2-m Himalayan Chandra Telescope (HCT)
of the Indian Astronomical Observatory (IAO{\footnote{\url{www.iiap.res.in/centers/iao}}}), Hanle, India. 
HFOSC is equipped with a 2K $\times$ 4K pixels, SITe CCD chip. The central 
2K $\times$ 2K pixels of the chip was used for imaging, which 
corresponds to a field of view of 10 $\times$ 10 arcmin$^2$ at an image scale of 0.296 arcsec pixel$^{-1}$. 
The gain and readout noise of the CCD camera are 1.22 electrons\,ADU$^{-1}$ and 
4.87 electrons, respectively. 
Photometric monitoring of SN 2012dn in the Bessell's $U$,  $B$, $V$, $R$ and $I$ filters,  
available with the HFOSC began on 2012 July 12 (JD 2456121.28), shortly after
discovery. The observations were continued till 2012 November 10 (JD 2456242.05), 
until the supernova  went behind the Sun. Photometric standard star fields from Landolt \citep{lan92}  were observed during photometric nights on 2012 July 16, 23 and October 13 for 
calibration of a sequence of secondary standard stars in the field of SN 2012dn.
Several bias images were taken on each night during the observation. Flat-field 
sky images in the $UBVRI$ bands were obtained during evening and morning 
twilight hours.

The data were processed in the standard manner, using various tasks available 
within the Image Reduction and Analysis Facility ({\sc iraf}{\footnote {{\sc 
iraf} is distributed 
by the National Optical Astronomy Observatories, which are operated by the 
Association of Universities for Research in Astronomy, Inc., under cooperative 
agreement with the National Science Foundation}}) software package.  The data 
were bias corrected with a master bias frame, which is  median of all the bias 
frames taken throughout the night. Median combined normalized flat-field images of the twilight 
sky were used for  flat field corrections. After this, cosmic ray hits were removed. 

\begin{table*}
\caption{Magnitudes of secondary standard stars in the field of SN 2012dn. The stars are marked in Fig. \ref{fig_std}.}
\begin{tabular}{lccccc}
\hline\hline
ID & U  & B & V &  R & I \\
\hline\hline     
1  &15.039 $\pm$ 0.007&14.949 $\pm$ 0.017&	14.313 $\pm$ 0.014&	13.982 $\pm$ 0.011&	13.578 $\pm$ 0.005\\ 
2  &15.092 $\pm$ 0.007&15.031 $\pm$ 0.023&	14.433 $\pm$ 0.015&	14.110 $\pm$ 0.011&	13.711 $\pm$ 0.007\\
3  &16.431 $\pm$ 0.008&15.674 $\pm$ 0.010&	14.779 $\pm$ 0.018&	14.310 $\pm$ 0.011&	13.809 $\pm$ 0.011\\
4  &17.766 $\pm$ 0.014&16.371 $\pm$ 0.023&	15.088 $\pm$ 0.008&	14.339 $\pm$ 0.011&	13.614 $\pm$ 0.011\\
5  &16.995 $\pm$ 0.016&15.970 $\pm$ 0.023&	14.957 $\pm$ 0.019&	14.455 $\pm$ 0.012&	13.948 $\pm$ 0.005\\
6  &16.160 $\pm$ 0.008&15.785 $\pm$ 0.014&	14.930 $\pm$ 0.016&	14.484 $\pm$ 0.004&	13.926 $\pm$ 0.022\\
7  &15.610 $\pm$ 0.021&15.692 $\pm$ 0.023&	15.229 $\pm$ 0.024&	14.978 $\pm$ 0.006&	14.630 $\pm$ 0.021\\
8  &16.173 $\pm$ 0.012&16.280 $\pm$ 0.017&	15.757 $\pm$ 0.015&	15.467 $\pm$ 0.010&	15.116 $\pm$ 0.010\\
\hline
\end{tabular}
\label{tab_std}
\end{table*}

\begin{table*}
\caption{Optical $UBVRI$ photometry of SN 2012dn with HCT.}
\begin{tabular}{lcccccccc}
\hline\hline
 Date & J.D. & Phase\rlap{*} & U & B & V & R & I\\
      &      & (days)        &   &   &   &   &  \\
\hline\hline
12/07/2012&2456121.28&$-$11.61 & 14.863 $\pm$ 0.030&  15.447 $\pm$ 0.013&  15.378 $\pm$ 0.011&  15.271 $\pm$ 0.006&  15.182 $\pm$ 0.028\\	
14/07/2012&2456123.30&$-$09.59 & 14.383 $\pm$ 0.038&  15.021 $\pm$ 0.024&  14.962 $\pm$ 0.011&  14.892 $\pm$ 0.026&  14.835 $\pm$ 0.027\\	
16/07/2012&2456125.28&$-$07.61 & 14.257 $\pm$ 0.060&  14.747 $\pm$ 0.008&  14.706 $\pm$ 0.015&  14.667 $\pm$ 0.016&  14.640 $\pm$ 0.010\\	
17/07/2012&2456126.31&$-$06.58 & 14.071 $\pm$ 0.024&  14.660 $\pm$ 0.020&  14.602 $\pm$ 0.014&  14.574 $\pm$ 0.019&  14.578 $\pm$ 0.026\\	
19/07/2012&2456128.30&$-$04.59 & 13.969 $\pm$ 0.033&  14.498 $\pm$ 0.017&  14.443 $\pm$ 0.020&  14.436 $\pm$ 0.019&  14.438 $\pm$ 0.026\\	
23/07/2012&2456132.26&$-$00.62 & 13.954 $\pm$ 0.027&  14.386 $\pm$ 0.013&  14.312 $\pm$ 0.013&  14.285 $\pm$ 0.015&  14.343 $\pm$ 0.010\\	
27/07/2012&2456136.30& 03.41 & 14.070 $\pm$ 0.037&  14.429 $\pm$ 0.010&  14.306 $\pm$ 0.006&  14.248 $\pm$ 0.013&  14.307 $\pm$ 0.014\\     
29/07/2012&2456138.26& 05.37 &                   &  14.499 $\pm$ 0.024&  14.338 $\pm$ 0.028&  14.274 $\pm$ 0.025&  14.318 $\pm$ 0.022\\	
03/08/2012&2456143.24& 10.35 & 14.567 $\pm$ 0.178&  14.862 $\pm$ 0.038&  14.425 $\pm$ 0.019&  14.356 $\pm$ 0.028&  14.320 $\pm$ 0.012\\	
19/08/2012&2456159.22& 26.33 &                   &  16.490 $\pm$ 0.015&  15.202 $\pm$ 0.007&  14.823 $\pm$ 0.013&  14.397 $\pm$ 0.021\\	
20/08/2012&2456160.16& 27.27 &                   &  16.598 $\pm$ 0.028&  15.259 $\pm$ 0.022&  14.843 $\pm$ 0.011&  14.465 $\pm$ 0.014\\	
14/09/2012&2456185.18& 52.29 &                   &  17.485 $\pm$ 0.012&  16.199 $\pm$ 0.015&  15.836 $\pm$ 0.016&  15.359 $\pm$ 0.009\\	
15/09/2012&2456186.13& 53.24 & 17.465 $\pm$ 0.044&  17.505 $\pm$ 0.018&  16.257 $\pm$ 0.010&  15.884 $\pm$ 0.014&  15.398 $\pm$ 0.015\\	
16/09/2012&2456187.13& 54.24 & 17.522 $\pm$ 0.026&  17.546 $\pm$ 0.017&  16.283 $\pm$ 0.019&  15.915 $\pm$ 0.014&  15.415 $\pm$ 0.009\\	
24/09/2012&2456195.07& 62.18 &                   &  17.823 $\pm$ 0.023&  16.573 $\pm$ 0.019&  16.261 $\pm$ 0.013&  15.794 $\pm$ 0.024\\ 
25/09/2012&2456196.06& 63.17 & 17.849 $\pm$ 0.047&  17.815 $\pm$ 0.064&  16.596 $\pm$ 0.018&  16.301 $\pm$ 0.031&  15.864 $\pm$ 0.035\\	
30/09/2012&2456201.10& 68.21 &                   &  17.964 $\pm$ 0.030&  16.806 $\pm$ 0.013&  16.521 $\pm$ 0.013&  16.085 $\pm$ 0.015\\	
01/10/2012&2456202.07& 69.18 &                   &  17.962 $\pm$ 0.026&  16.815 $\pm$ 0.022&  16.533 $\pm$ 0.020&  16.142 $\pm$ 0.008\\	
02/10/2012&2456203.06& 70.17 & 18.043 $\pm$ 0.042&  17.985 $\pm$ 0.014&  16.846 $\pm$ 0.015&  16.553 $\pm$ 0.012&  16.160 $\pm$ 0.016\\	
13/10/2012&2456214.06& 81.17 &                   &  18.281 $\pm$ 0.016&  17.306 $\pm$ 0.011&  17.102 $\pm$ 0.013&  16.704 $\pm$ 0.017\\	
17/10/2012&2456218.04& 85.15 &                   &  18.422 $\pm$ 0.014&  17.451 $\pm$ 0.016&  17.256 $\pm$ 0.017&  16.886 $\pm$ 0.011\\	
19/10/2012&2456220.08& 87.20 &                   &  18.495 $\pm$ 0.014&  17.540 $\pm$ 0.012&  17.343 $\pm$ 0.011&  16.987 $\pm$ 0.017\\	
27/10/2012&2456228.04& 95.15 &                   &  18.687 $\pm$ 0.028&  17.860 $\pm$ 0.016&  17.707 $\pm$ 0.016&  17.351 $\pm$ 0.022\\	
30/10/2012&2456231.05& 98.16 &                   &  18.742 $\pm$ 0.030&  17.939 $\pm$ 0.020&  17.792 $\pm$ 0.022&  17.474 $\pm$ 0.016\\	
10/11/2012&2456242.05&109.16 &                   &  18.968 $\pm$ 0.074&  18.342 $\pm$ 0.027&  18.200 $\pm$ 0.021&  17.976 $\pm$ 0.026\\	   
\hline                         
\multicolumn{8}{l}{\rlap{*}\ ~Observed phase with respect to the epoch of $B$ band maximum: JD = 2456132.89.}
\end{tabular}		            
\label{tab_snmag}	       
\end{table*}

Aperture photometry was performed on the standard stars with {\sc daophot} package of {\sc iraf} at an optimal aperture 
determined by the aperture growth curve. Using the bright stars in the field, an aperture correction was estimated by measuring the magnitude difference at optimal aperture
and at an aperture close to full width half maximum (FWHM) of stellar profile. This correction was applied to the magnitude obtained at smaller aperture. Instrumental magnitudes obtained this way were corrected for atmospheric  extinction using the average value of extinction 
coefficients for the site \citep{sta08}. The average colour terms for the system were used 
to determine the photometric zero points on photometric nights. These zero 
points and average colour terms were then used to calibrate a sequence of 
secondary standards in the supernova field observed on the same nights as 
the standard fields. Calibrated $UBVRI$ magnitudes of the secondary standards, 
averaged over 3 nights are given in Table \ref{tab_std} and have been marked 
in Fig. \ref{fig_std}. These magnitudes were used to calibrate the supernova
magnitude  obtained on  other nights.

The magnitudes of supernova and the secondary standards in the supernova 
frames were obtained using point spread function (PSF) fitting technique, 
with a fitting radius equal to the FWHM of the stellar profile. 
The magnitudes of secondary standards were also estimated using aperture 
photometry. Magnitude difference between the aperture and profile fitting 
techniques was obtained using bright secondary standards. This correction 
was applied to the supernova magnitude. Zero-points on each night were 
determined using the secondary standards in the supernova field and finally 
supernova magnitudes were calibrated differentially with respect to the 
secondary standards.  The final calibrated supernova magnitudes and errors 
in  $U$,  $B$, $V$, $R$ and $I$ bands are listed in Table \ref{tab_snmag}.

\subsubsection{UV-Optical Photometry using $Swift$ UVOT}
Supernova SN 2012dn was monitored with the Ultra Violet Optical Telescope (UVOT)
on board the {\it Swift} satellite during 2012 July 13 (JD 2456121.57) to 
August 11 (JD 2456151.30). The observations were made in the 3 broad-band
UV filters, ($uvw2$ : 1928 \AA, $uvm2$ : 2246 \AA, $uvw1$ : 2600 \AA) and 
three broad-band optical filters ($u$ : 3465 \AA, $b$ : 4392 \AA \ $v$ : 5468 \AA).
The {\it Swift} UVOT data of SN 2012dn available at {\it Swift} archive are used in this study.
The data reduction was carried out using various packages in the {\sc heasoft} (the High
 Energy Astrophysics Software) following methods of \citet{poo08} and \citet{bro09}.
Aperture photometry was performed using {\it uvotsource} program to extract apparent magnitude of the supernova. During the early phase when supernova was bright  
an aperture size of 5 arcsec was used while  in the  late phase smaller aperture 
of 3.5 arcsec was used to estimate supernova magnitude and  aperture correction as 
listed by \citet{poo08} was applied. Sky in the nearby region was estimated using an aperture as was used for the supernova.  The {\it Swift} UVOT magnitudes of 
SN 2012dn is listed in Table \ref{tab_snmag_uvot}. 
\begin{table*}
\caption{UV-Optical photometry of SN 2012dn with {\it Swift} UVOT.}
\begin{tabular}{lcccccccc}
\hline\hline
 Date & JD & Phase\rlap{*} & $uvw2$ & $uvm2$ & $uvw1$ & $u$ & $b$ & $v$\\
     &   & (days)    &   &   &   &   & & \\
\hline\hline
13/07/2012&2456121.57 &$-$11.32 & 16.22$\pm$0.06& &15.24$\pm$0.05&14.37$\pm$0.04 &15.35$\pm$0.04& \\
16/07/2012&2456125.10 &$-$7.79 & 15.91$\pm$0.06&15.71$\pm$0.06&14.75$\pm$0.05&13.87$\pm$0.04&14.73$\pm$0.04&14.66$\pm$0.06\\
18/07/2012&2456126.51 &$-$6.38 & 15.91$\pm$0.05&15.82$\pm$0.05&14.71$\pm$0.05&13.73$\pm$0.04&14.59$\pm$0.04&14.59$\pm$0.05\\
20/07/2012&2456128.92 &$-$3.97 & 15.98$\pm$0.06&15.90$\pm$0.06&14.74$\pm$0.05&13.71$\pm$0.04&14.44$\pm$0.04&14.49$\pm$0.05\\
24/07/2012&2456132.53 &$-$0.36 & 16.22$\pm$0.07&16.35$\pm$0.07&14.92$\pm$0.05&13.80$\pm$0.04&14.38$\pm$0.04&14.26$\pm$0.05\\
26/07/2012&2456134.53 &1.64 & 16.57$\pm$0.07&16.47$\pm$0.07&15.18$\pm$0.05&13.92$\pm$0.04&14.41$\pm$0.04&14.29$\pm$0.05\\
30/07/2012&2456138.79 &5.90 & 16.73$\pm$0.06&16.98$\pm$0.12&15.57$\pm$0.05&14.59$\pm$0.04&14.50$\pm$0.04&14.34$\pm$0.04\\
05/08/2012&2456144.55 &11.66 & 17.31$\pm$0.10&17.47$\pm$0.13&16.27$\pm$0.08&14.96$\pm$0.05&14.91$\pm$0.04&14.44$\pm$0.05\\
11/08/2012&2456151.30 &18.41 & 17.91$\pm$0.14&18.06$\pm$0.23&16.96$\pm$0.10&15.79$\pm$0.06&15.66$\pm$0.05&14.82$\pm$0.05\\
\hline
\multicolumn{8}{l}{\rlap{*}\ ~Observed phase with respect to the epoch of $B$ band maximum: JD = 2456132.89.}
\end{tabular}			    
\label{tab_snmag_uvot}	    
\end{table*}

\subsection{Spectroscopy}
\label{sec:spec_observation}
Spectroscopic observations of SN 2012dn were made in Gr\#7 (wavelength range 
3500--7800 \AA) and Gr\#8 (wavelength range 5200--9250 \AA) using HFOSC at a 
spectral resolution of $\sim$ 7 \AA. Sixteen spectra, covering the phase 
$-$11.58 d (JD 2456121.31) to +98.18 d (JD 2456231.07) relative to the $B$ 
band maximum were obtained. A journal of spectroscopic observation is 
given in Table \ref{tab_spec}. Arc lamp spectra of FeNe (in Gr\#8) and FeAr 
(in Gr\#7) were obtained for wavelength calibration. Spectrophotometric 
standard stars were observed for flux calibration. Data reduction was done in 
the standard manner using various tasks within {\sc iraf}.  One-dimensional 
spectra were extracted using the optimal extraction method. The extracted 
spectra were wavelength calibrated using the arc lamp spectra. The instrumental 
response was removed from the wavelength calibrated spectra using spectra of  
spectrophotometric standard stars,  preferably observed on the same night. In 
case of non-availability of standard star observation on a particular night, the
standard stars observed during nearby nights were used. The spectra obtained in 
two wavelength regions were combined by scaling to a weighted mean to get the 
final spectrum on a relative flux scale. The spectra were then brought to an 
absolute flux scale by applying zero point corrections obtained from photometry.
A redshift correction of \textit{z} = 0.010 (adopted from NED) and total 
reddening correction of $E(B-V)$ = 0.18 mag (refer section \ref{sec:reddening}) 
were applied to the final supernova spectra. The telluric lines have not been 
removed from the spectra.         

\begin{table}
\caption{Log of spectroscopic observations of SN 2012dn.}
\begin{tabular}{lccc}
\hline\hline
Date & J.D. & Phase\rlap{*} & Range \\
     &      & days & \AA  \\
\hline\hline
12/07/2012&   2456121.31 &$-$11.58  & 3500-7800; 5200-9250\\
16/07/2012&   2456125.29 &$-$7.60  & 3500-7800; 5200-9250\\
18/07/2012&   2456127.29 &$-$5.60  & 3500-7800; 5200-9250\\
23/07/2012&   2456132.29 &$-$0.60  & 3500-7800; 5200-9250\\
28/07/2012&   2456137.30 &  4.41  & 3500-7800; 5200-9250\\
29/07/2012&   2456138.28 & 5.39   & 3500-7800; 5200-9250\\
03/08/2012&   2456143.26 & 10.37  & 3500-7800; 5200-9250\\
19/08/2012&   2456159.23 & 26.34  & 3500-7800          \\
14/09/2012&   2456185.21 & 52.32  & 3500-7800; 5200-9250\\
15/09/2012&   2456186.18 & 53.29  & 3500-7800; 5200-9250\\
16/09/2012&   2456187.15 & 54.26  & 3500-7800          \\
24/09/2012&   2456195.10 & 62.21  & 3500-7800; 5200-9250\\
01/10/2012&   2456202.08 & 69.19  & 3500-7800; 5200-9250\\
13/10/2012&   2456214.10 & 81.21  & 3500-7800; 5200-9250\\
19/10/2012&   2456220.10 & 87.21  & 3500-7800; 5200-9250\\
30/10/2012&   2456231.07 & 98.18  & 3500-7800; 5200-9250\\
\hline
\multicolumn{4}{l}{\rlap{*}\ ~Relative to the epoch of $B$ band maximum.}
\end{tabular}
\label{tab_spec}
\end{table}

\section{Light curve analysis}
\label{sec:light_curve_analysis}
\subsection{Light curves and colour curves}
\label{sec:light_curve}
\begin{figure}
\resizebox{\hsize}{!}{\includegraphics{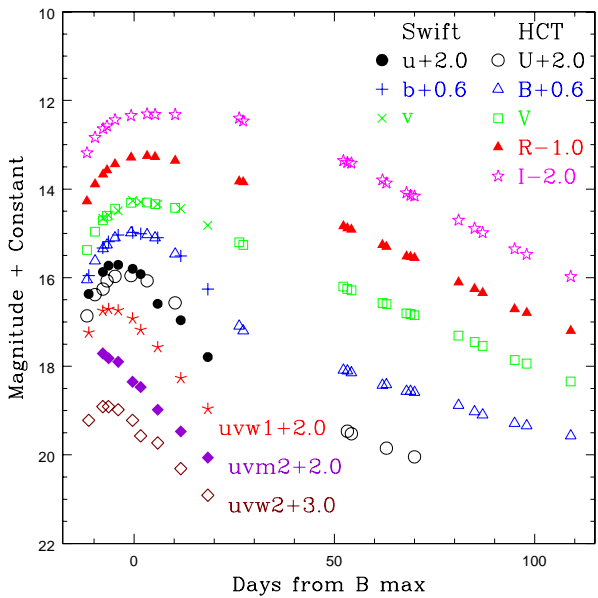}}
\caption[]{$UBVRI$ and {\it Swift} UVOT  light curves of SN 2012dn. The light curves have been shifted by the 
amount indicated in legend. The phase is measured in days from the $B$ band maximum. (A colour version of this figure is available in the online journal.)}
\label{fig_light}
\end{figure}

The light curves of SN 2012dn in the $U$,  $B$, $V$, $R$ and $I$ bands are 
plotted in Fig. \ref{fig_light}. The UVOT magnitudes are also plotted
in the same figure. There is a good agreement between  the UVOT magnitudes in $b$ and $v$ 
bands  and the $B$ and $V$ magnitudes obtained with the HCT. The observed data set is
 used to derive various  photometric parameters of SN 2012dn, which are tabulated in Table 
\ref{tab_parameter}. The date of maximum light and brightness at maximum light  
were estimated by fitting a cubic spline to the points around maximum. The $B$ 
band light curve reached a maximum brightness of 14.38 $\pm$ 0.01 mag on 
JD 2456132.89 $\pm$ 0.19. The post-maximum decline rate during the first 15 
days, $\Delta m_{15}(B)$, is estimated to be 0.90 $\pm$ 0.04. The reddening 
corrected decline rate parameter ($\Delta m_{15}(B)_\text{true}$), estimated 
using the relation by \citet{phi99} is 0.92. The $\Delta m_{15}(B)$ of SN 2012dn
is comparable  to those of SN 1991T \citep{lir98}, SNF 20080723-012 \citep{sca12}, 
slower  than  normal SNe Ia SN 2003du \citep*{anu05}, SN 2005cf \citep{pas07}
and faster than the super-Chandra SNe Ia SN 2006gz \citep{hic07}, SN 2007if 
\citep{sca10}, and SN 2009dc \citep{yam09,sil11,tau11}. 

\begin{table*}
\caption{Photometric parameters of SN 2012dn.}
\begin{tabular}{lcccccc}
\hline\hline
Band & JD (Max)  & $m_{\lambda}^\text{max}$& $M_{\lambda}^\text{max}$&  $\Delta m_{15}(\lambda)$ &Decline rate\rlap{*} &Colours at $B$ max\\
     &   & & & & during (50--100 d) & \\
\hline\hline  
$U$ & 2456130.42 $\pm$ 0.06  &13.936 $\pm$ 0.030  & $-$20.094 $\pm$ 0.155 &0.80 $\pm$ 0.04&3.786&\\
$B$ & 2456132.89 $\pm$ 0.19  &14.384 $\pm$ 0.017  & $-$19.516 $\pm$ 0.152 &0.90 $\pm$ 0.04&2.764& $(U-B)_0$ = $-$0.620 $\pm$ 0.002 \\
$V$ & 2456134.52 $\pm$ 0.10  &14.295 $\pm$ 0.016  & $-$19.417 $\pm$ 0.152 &0.36 $\pm$ 0.02&3.824& $(B-V)_0$ = $-$0.100 $\pm$ 0.002 \\
$R$ & 2456135.46 $\pm$ 0.06  &14.245 $\pm$ 0.018  & $-$19.328 $\pm$ 0.152 &0.31 $\pm$ 0.03&4.327& $(V-R)_0$ = $-$0.044 $\pm$ 0.002\\
$I$ & 2456136.16 $\pm$ 0.17  &14.307 $\pm$ 0.020  & $-$19.114 $\pm$ 0.153 &               &4.674& $(R-I)_0$ = $-$0.195 $\pm$ 0.003 \\
\hline
\multicolumn{7}{l}{\rlap{*}\  ~in unit of mag (100 d)$^{-1}$ and epoch is relative to $B$ band maximum.}
\end{tabular}
\label{tab_parameter}
\end{table*}
The maximum in $U$ band (13.936  $\pm$ 0.030 mag) occurred on JD 2456130.42 $\pm$ 0.06, 
$\sim$ 2.5 
d before the $B$ band maximum, while the supernova reached   maximum  in $V$, $R$ and $I$ bands 
at $\sim$  1.6 d (JD 2456134.52 $\pm$ 0.10), $\sim$ 2.6 d (JD 2456135.46 $\pm$ 0.06) and 
$\sim$ 3.3 d (JD 2456136.16 $\pm$ 0.17) after $B$ band maximum, respectively. In a recent study using the 
{\it Swift} UVOT data, \citet{brown14} have reported    $\Delta m_{15}(B)$ as 1.08$\pm$0.03. The  date of maximum and magnitude at maximum estimated in this work  are consistent with those reported by \citet{brown14}.   The difference in the value of $\Delta m_{15}(B)$ may be due 
to the methods adopted for its estimation. \citet{brown14} have determined it 
by stretching a template light curve \citep*[from MLCS2k2:][]{jha07} to the data between 
2 days before and 15 days after maximum light and interpolating from the 
stretched template,  whereas we have determined it by fitting a cubic spline to the 
observed data points around maximum.   

In normal and luminous SNe Ia, the maximum in $I$ band always precedes that of $B$ band. 
For example, in SN 1991T, $I$ band light curve reached  maximum  at $-$0.4 d with respect 
to $B$ band maximum, in SN 1990N the $I$ maximum occurred  at  $-$2.0 d \citep{lir98} 
and in  SN 2003du the maximum in $I$ band occurred at  $-$1.9 d \citep{anu05}.  
However,  in SN 2012dn the $I$ maximum  occurred  3.3 days  after 
 $B$ band maximum.  For SN 2012dn,  the delay in $I$ band maximum 
with respect to $B$ band is not in accordance  with the observed trend for normal 
and luminous SNe Ia. A similar behaviour was also observed in SN 2009dc \citep{tau11}.
Delayed $I$ band maximum with respect to $B$ band maximum has been observed 
in subluminous SN 1991bg-like objects \citep{tau08} and the peculiar subluminous 
supernova  SN 2005hk \citep{sah08}. 
\begin{figure}
\resizebox{\hsize}{!}{\includegraphics{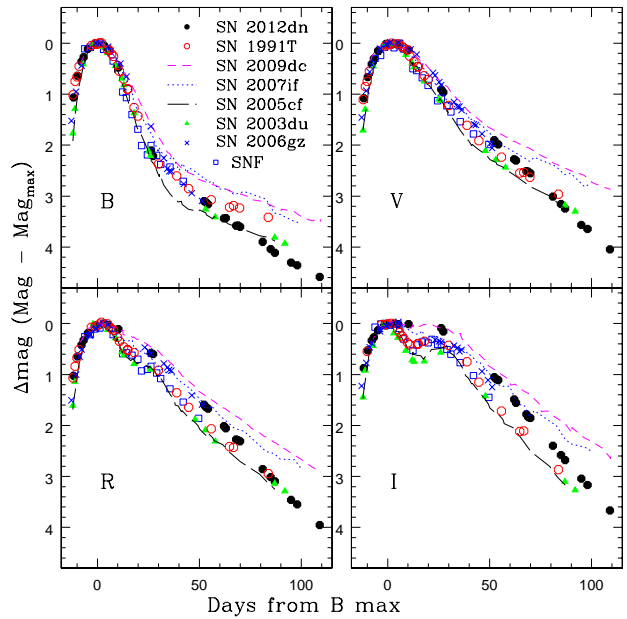}}
\caption[]{ $BVRI$ light curves of SN 2012dn compared with those of SN 1991T, SN 2009dc, SN 2007if, SN 2005cf, SN 2003du, SN 2006gz and SNF20080723-012. All the light curves have been shifted to match with their peak magnitudes and to the epoch of $B$ band maximum. (A colour version of this figure is available in the online journal.)}
\label{fig_lcomp}
\end{figure}

In Fig. \ref{fig_lcomp}, the $BVRI$ light curves of SN 2012dn are compared with those of 
the well studied normal  SNe Ia: 
SN 2003du ($\Delta m_{15}(B)$ = 1.04; \citealt{anu05}), 
SN 2005cf ($\Delta m_{15}(B)$ = 1.12; \citealt{pas07}), 
the luminous SNe Ia: SN 1991T ($\Delta m_{15}(B)$ = 0.95; \citealt{lir98}), 
SNF20080723-012 ($\Delta m_{15}(B)$ = 0.93; \citealt{sca12})
and super-Chandra SNe Ia: 
SN 2009dc ($\Delta m_{15}(B)$ = 0.71; \citealt{tau11}), 
SN 2007if ($\Delta m_{15}(B)$ = 0.71; \citealt{sca10}), 
SN 2006gz ($\Delta m_{15}(B)$ = 0.69; \citealt{hic07}). 
All the light curves in each band have been shifted to match their peak magnitudes 
and to the epoch of $B$ band maximum. 
From Fig. \ref{fig_lcomp}, we find that the light curve width of SN 2012dn is intermediate between normal and  super-Chandra SNe Ia.  In the early phase, the
 $B$ band light curve of SN 2012dn is similar to that of SN 1991T, however, during 
 late phase, SN 2012dn declines faster than  SN 1991T. A similar trend is seen in the $V$ band also. 

\begin{figure}
\resizebox{\hsize}{!}{\includegraphics{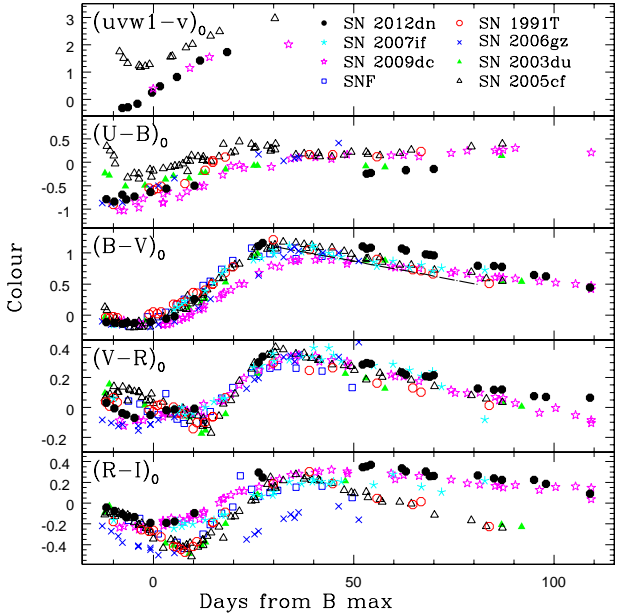}}
\caption[]{The reddening corrected $(uvw1-v)$, $(U-B)$, $(B-V)$, $(V-R)$ and $(R-I)$ colour curves of SN 2012dn 
plotted  with those of SN 1991T, SN 2003du, SN 2005cf, SN 2006gz, SN 2007if, SN 2009dc and SNF 20080723-012.
The dashed line drawn with the $(B-V)$ colour curve represents the Lira-Phillips relation \citep{phi99}. (A colour version of this figure is available in the online journal.)} 
\label{fig_colourcomp}
\end{figure}
The reddening corrected $(uvw1-v)$, $(U-B)$, $(B-V)$, $(V-R)$ and $(R-I)$ colour curves of 
SN 20012dn are plotted in Fig. \ref{fig_colourcomp} and their values at the time of $B$ band maximum are listed in Table \ref{tab_parameter}. Colour curves of some well studied SNe Ia (used for comparing the light curves) have also been plotted in the same figure for comparison. 
Following extinction law of \citet*{car89}, all the colour curves have been dereddened. The total reddening of 
$E(B-V)_{Gal+host}$ = 0.18 was used for SN 2012dn (refer section \ref{sec:reddening}) and for other SNe the reddening values were taken from their respective references.   

The $(U-B)$ colour of SN 2012dn is bluer as compared to the 
normal type Ia SNe SN 2003du and SN 2005cf. In the 
pre-maximum phase $(U-B)$ colour evolution of SN 2012dn is similar to  
SN 2006gz and marginally redder than SN 2009dc, but during the post-maximum phase 
it is significantly bluer. \citet{brown14} have noted  that during pre-maximum phase
SN 2012dn has bluer $(uvw1-v)$ colour as compared to other spectroscopically normal SNe  Ia
  \citep[refer Fig 3. of][]{brown14}. In spectroscopically normal SNe Ia
 the $(uvw1-v)$ and $(U-B)$ colours  evolve from red to blue, reaching a 
minimum  colour a few days before the optical maximum and then becoming redder again.  
The $(U-B)$ and $(uvw1-v)$ colour evolution of SN 2012dn donot follow this trend of
normal SNe Ia.    

From pre-maximum phase to 30 d after $B$ band maximum, the $(B-V)$ colour 
evolution of SN 2012dn is similar to the normal type Ia SNe 2003du, 2005cf, 
overluminous SN 1991T and super-Chandra SN 2007if. During 50--80 d after
$B$ maximum  SN 2012dn shows redder $(B-V)$ colour and  $\sim$ 100 d after 
$B$ band maximum $(B-V)$ colour of SN 2012dn
comes close to normal type Ia SNe  as well as super-Chandra SNe.
 The $(V-R)$ colour evolution of SN 2012dn follows those of normal and super-Chandra SNe. 
The  $(R-I)$ colour evolution of  SN 2012dn closely follows 
SN 2009dc and SN 2007if  and  is different from  that of SN 1991T and other
normal SNe Ia, which may be due to difference in the strength of 
the  $I$ band secondary maximum in these events.

\subsection{Luminous secondary maximum in $R$ and $I$ bands}
\begin{figure}
\resizebox{\hsize}{!}{\includegraphics*[trim = 0.0in 0mm 0in 3.7in, clip]{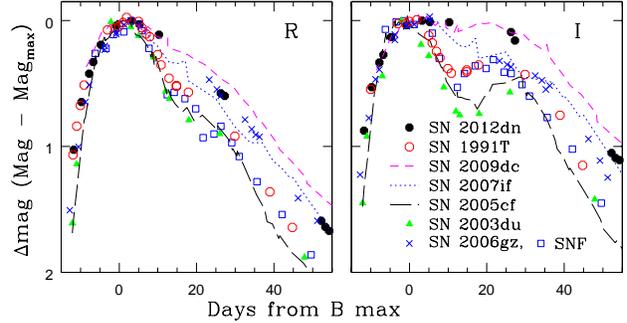}}
\caption[]{Same as Fig. \ref{fig_lcomp} but only for $R$ and $I$ bands during  early phase. The $I$ band peak of SN 2012dn has a  plateau-like feature. (A colour version of this figure is available in the online journal.)}
\label{fig_lcomp2}
\end{figure}

The $I$ band light curve of normal and SN 1991T-like SNe Ia shows 
secondary maximum around 21 to 30 d after $B$ band maximum \citep{lei00}, with  a pronounced 
minima in between the peaks. Around this epoch, a shoulder/rebrightening is also seen in 
the $R$ band light curve. In Fig. \ref{fig_lcomp2}, we  have compared early phase  $R$ 
and $I$ band light curves of SN 2012dn   with the light curves of SNe  used for comparison 
in Fig. \ref{fig_lcomp}. Our imaging data  does not have very good  coverage during early 
post-maximum phase. However, it is evident from Fig. \ref{fig_lcomp2} that the minima 
between the primary and secondary peak is absent/very weak in the  $I$ band light curve 
of SN 2012dn,  different from those of normal SNe Ia and SN 1991T-like objects. 
In SN 2009dc, the minima between primary and secondary maxima was not so pronounced and 
the  early post-maximum $R$ and $I$ light curves flatten to give a plateau-like 
appearance \citep{sil11,tau11}. 
   
It is apparent from  Fig. \ref{fig_lcomp2} that in the $I$ band, normal SNe Ia 
 SN 2003du and SN 2005cf show deep  minimum  between primary and secondary maxima. The  
depth of minimum with respect to primary maximum 
is $\sim$ 0.7 mag for normal SNe Ia (used here) while  in SN 1991T it  is  $\sim$ 0.4 mag. 
As compared to the primary  maximum, the secondary maximum is fainter by $\sim$ 0.5 mag 
and $\sim$ 0.3 mag in  normal  and  SN 1991T-like luminous events, respectively.
 In SN 2009dc, the primary and secondary maxima are almost equally bright and minimum  
seems to be unnoticeable. This gives a broader, plateau-like appearance to the $I$ band 
light curve around maximum light.   The $I$ band light curve of  SN 2006gz 
and SN 2007if  appears to have linear decline  during the early post-maximum phase.   
 From Fig. \ref{fig_lcomp2}, it is clear that in the early post-maximum phase, $I$ band 
light curve of SN 2012dn appears to be very similar to that of SN 2009dc. The strength of 
secondary maximum in both these objects is similar. Light curve of SN 2012dn in $R$ band 
during the early post-maximum decline is  similar to SN 2006gz and SN 2007if,  showing a 
stronger bump  compared to normal SNe Ia, but weaker than SN 2009dc. 

Early phase post-maximum ($<$ 50 d) decline rates  ($\Delta m_{50}(\lambda)$) for SN 2012dn and other SNe in comparison have been calculated in the $BVRI$ bands and are listed in Table \ref{tab_decline}. From this analysis and from Fig. \ref{fig_lcomp2}, it is clear that the sequence of rebrightening/broadening of $R$ and $I$ band light curves of SNe in comparison is: normal (SN 2003du, SN 2005cf) $\rightarrow$ luminous (SNF, SN 1991T) $\rightarrow$ super-Chandra (SN 2006gz, SN 2007if) $\rightarrow$ (SN 2012dn) $\rightarrow$ super-Chandra (SN 2009dc). SN 2009dc appears to have the most rebrightening/broadening during this epoch.
\begin{table*}
\caption{Decline rate of SN 2012dn compared with other SNe Ia.}
\begin{tabular}{lccccccccc}
\hline\hline
SNe & $U$ & $B$  & $V$ & $R$ & $I$  & $\Delta m_{50}(B)$ & $\Delta m_{50}(V)$ & $\Delta m_{50}(R)$ & $\Delta m_{50}(I)$ \\
    & \multicolumn{5}{c}{\rlap{*}\, Decline rate during ~50--100 d} & & \\
\hline\hline  
SN 2012dn &3.786 &2.764  &3.824  &4.327  &4.674  &3.05&1.88& 1.62& 1.11 \\
SN 2009dc &1.950 &1.452  &2.094  &2.703  &3.034  &2.66&1.68& 1.38& 0.91  \\
SN 2007if &   -  &1.863  &1.918  &2.740  &3.104  &-&-&  -  &  - \\    
SN 1991T  &1.831 &1.399  &2.718  &3.335  &4.346  &2.98&2.14& 1.89& 1.37  \\
SN 2005cf &2.198 &1.580  &2.581  &3.184  &4.132  &3.29&2.31& 2.08& 1.50  \\
SN 2003du &2.363 &1.652  &2.649  &3.110  &3.979  &3.18&2.22& 1.98& 1.46  \\
\hline
\multicolumn{8}{l}{\rlap{*}\  ~in unit of mag (100 d)$^{-1}$ and epoch is relative to $B$ band maximum.}
\end{tabular}
\label{tab_decline}
\end{table*}

The secondary maximum in the $I$ through $K$ band light curves are thought to 
be associated with the ionization state of the ejecta \citep{pin00,liw03,kas06}. Based on numerical simulations, \citet{kas06} has 
shown that the secondary maximum of SNe Ia is directly related to the 
ionization evolution of IGEs in the ejecta. It is identified with the onset and 
recession of the 2 $\rightarrow$ 1 (transition from doubly to singly ionized 
states) ionization front into iron-rich ejecta. Other factors such as change 
in flux mean opacity, mass of $^{56}$Ni and mixing of $^{56}$Ni into the ejecta
affect the occurrence and strength of secondary maxima. Mixing of $^{56}$Ni outward into the 
region of IMEs advances the secondary maximum, causing it to merge with the 
first maximum. This also results in broadening the peak in $R$ and $I$ band 
light curves, and the disappearance of the secondary maximum. The strength of 
the secondary maximum is found to be a good measure of the amount of IGEs 
synthesized in the explosion. 

In subluminous SNe Ia, because of the small amount of $^{56}$Ni, the secondary 
maximum occurs at early times and  merges with the first maximum, leading to an 
absence of the secondary maximum. In the case of the low luminosity, peculiar 
SN Ia SN 2002cx, the broad peak in the $R$ band and the nearly constant, plateau 
phase for $\sim$ 20 d in the $I$ band, seen around maximum, were suggested, by 
\citet{liw03}, to be a result of the rapid change in the ionization stage 
(from Fe\,{\sc iii} to Fe\,{\sc ii}). According to \citet{liw03}, this change in 
the ionization stage causes an early release of the residual stored energy, 
resulting in the merger of the primary and secondary peak, giving rise to a 
broadened peak in the $R$ band and a plateau phase in the $I$ band light curves.

The observed absolute magnitude of SN 2012dn clearly indicates the supernova is
not subluminous. Hence the apparent absence of a secondary maximum, is not due
to low production of $^{56}$Ni. As suggested earlier, the secondary maximum in
the $I$ band is luminous, giving it a broad, plateau appearance. This hints 
towards a larger production of IGEs. The large production of IGEs and their 
mixing to higher velocities in the explosion thus appears to be the cause of 
luminous secondary maximum in SN 2012dn.

\subsection{Enhanced fading in the late phase}
\begin{figure}
\resizebox{\hsize}{!}{\includegraphics*[trim = 0.0in 0mm 0in 3.7in, clip]{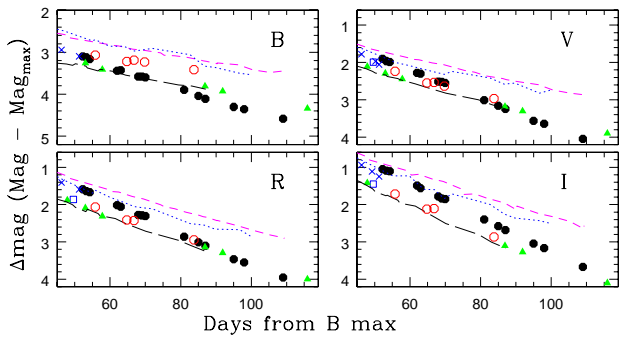}}
\caption[]{Same as Fig. \ref{fig_lcomp} but during  late phase. (A colour version of this figure is available in the online journal.)}
\label{fig_lcomp3}
\end{figure}

During late phase ($\sim$ 50--100 d after $B$ band maximum), evolution of light curves of 
SN 2012dn is different from other well studied events (refer Fig. \ref{fig_lcomp} and Fig. \ref{fig_lcomp3}).
The late phase decline rate of SN 2012dn and  other events  is  estimated by a least 
square fit to the observed data in $U$, $B$, $V$, $R$, $I$ bands and is listed in Table \ref{tab_decline}. 
The late phase  decline rate of SN 2012dn is faster than that of other objects in comparison. 

Steepening in the light curve has also been observed in the super-Chandra SNe Ia SN 2006gz and SN 2009dc, 
but  at much later phases. Observation of SN 2006gz a year after 
the $B$ band maximum revealed  that it  was fainter than the normal type Ia SN 2003du 
by a factor of $\sim$ 4.  
SN 2006gz faded dramatically sometime between the peak phase and the nebular 
phase \citep{mae09}. 
Likewise, SN 2009dc showed an increase in the decline rate $\sim$ +200 d,  and at $\sim$ +400 d, it was no longer luminous than SN 2003du \citep{tau13}.
Though the peak luminosity of  SNF20080723-012 was  similar to SN 2006gz and SN 1991T, 
it also showed signature of rapid decline and  $\sim$ 200 d after maximum it was 
fainter than SN 1991T.  Among the super-Chandra SNe Ia  SN 2007if showed the least 
steepening in the light curve. It had a nearly constant decline rate during  
$\sim$ 50--300 d.

Various mechanisms, such as positron escaping, infra-red catastrophe (IRC), dust formation and reprocessing of optical light to longer wavelengths, have been proposed to explain the 
observed departure in the late phase decline rate  from the expected one due to 
$^{56}$Co$\rightarrow$$^{56}$Fe decay.

Positrons are produced in the ejecta of SNe Ia by $^{56}$Co$\rightarrow$$^{56}$Fe decay.
Around  50 d after explosion, the energy deposition is dominated by the interaction 
involving decay products of the $^{56}$Co$\rightarrow$$^{56}$Fe chain, namely,  gamma-ray photons 
and positrons. Because of the significantly high energy available with  photons per decay
 than positrons, during the early phase,  energy deposition is dominated by photons. At later phase ($\sim$ 100 d after explosion),  the energy deposition is dominated by positrons.  At low 
ejecta density, positron may escape depending on the nature of the magnetic field 
\citep*{mil01}. The escape of positron causes reduction in the energy deposition 
rates to the ejecta, which in turn leads to fainter light curves. This effect 
is found to be very low even at very late phase;  at $\sim$ 400 d, it  can account
for at the most $\sim$ 0.5 mag drop in the light curve \citep{mae09}. 
Hence, the observed  early steepening ($\sim$ 60 d after $B$ band maximum) in the 
light curve of SN 2012dn  can not be a consequence of escape of  
positrons. \citet{tau13} have suggested  that the
observed luminosity drop is  probably the outcome of the flux distribution into
infra-red, which could be accomplished by an early infra-red catastrophe (IRC)
or by dust formation.  

The IRC  causes  shift of bulk emission from optical to infra-red wavelengths 
at late phases \citep{axe88}, resulting in  significant drop in the late time 
optical light curves. This occurs  when the density and temperature of the ejecta 
fall below a critical value during the expansion of ejecta.  However, there is no 
indication of occurrence of IRC in SNe Ia \citep[see also][]{mae09,tau13}.   

Dust formation in the ejecta of SNe Ia has been explored theoretically by \citet{noz11}. 
They have shown that similar to core-collapse SNe,  dust can form in the ejecta of  
SNe Ia also. Because of  low gas density in the ejecta of SNe Ia,  the gas 
temperature decreases rapidly, resulting 
in much earlier condensation of dust ($\sim$ 100--300 d) than those in 
type IIP SNe ($>$ 300 d). 

The pre-maximum spectra of SN 2012dn show signature of C\,{\sc ii} $\lambda$6580 feature 
(refer section \ref{sec:premax_spec}), indicating the presence of unburned C. \citet{noz11} 
have shown that the C present in the ejecta can lead to formation of graphite dust in the 
supernova ejecta. The unburned C in the ejecta may also combine with  O to form CO molecules affecting  the composition, size and mass of dust grain. Further, radioactivity due to 
$^{56}$Ni produced in the explosion, destroys the molecules making C free for dust formation. 
The slow expansion velocity of the  ejecta   results in comparatively  high densities at late 
phases, which may also promote formation of dust in the supernova ejecta.   

It thus appears that   the observed fast decline  in 
SN 2012dn might be due to dust formation.   The onset  of dust formation in the ejecta 
should also increase the luminosity in the near-infrared (NIR) bands. However, due to non-availability of the NIR data we are unable to  confirm.   
 For super-Chandra objects, the observed  differences in the onset  of steepening in the light curve and  magnitude drop  can be understood in terms of different time-scales and intensities of dust formation \citep{tau13}.
                                   
\subsection{Reddening and Absolute magnitudes}
\label{sec:reddening}
Dust map of \citet*{sch98} gives  colour excess of  $E(B-V)$ = 0.06 mag 
towards SN 2012dn due to the ISM within the Milky Way.  Narrow Na\,{\sc i} D absorption
 lines from the Milky Way and the host galaxy are seen in our medium resolution
 spectra of SN 2012dn obtained near maximum light, with an equivalent width of 
0.69 \AA\, and 0.75 \AA\,, respectively. The empirical relation between reddening and 
equivalent width of Na\,{\sc i} D line  \citep*{tur03} gives colour 
excess of $E(B-V)$ = 0.12 mag from the host. 
For normal SNe Ia, the extinction within the host
galaxy can also be  estimated using the Lira-Phillips relation \citep{phi99}. In Fig. \ref{fig_colourcomp} we have plotted the 
Lira-Phillips relation along with the  $(B-V)$ colour
 evolution during 30--90 d after  maximum light. The Lira-Phillips relation 
gives the colour excess within the host as $E(B-V)$ = 0.43 mag, which is much higher 
than the reddening value estimated using the Na\,{\sc i} D lines. Similar discrepancy 
between the reddening estimates from the Na\,{\sc i} D lines and Lira-Phillips relation is 
noticed for  SN 2009dc \citep{yam09} and SN 2007if \citep{sca10} also.
Further, the  $(B-V)$ colour evolution of SN 2012dn appears to be slower 
than that expected from the Lira-Phillips relation (refer Fig. \ref{fig_colourcomp}), 
indicating that Lira-Phillips relation may not hold for these
objects. 
 
Host galaxy reddening can be estimated using the observed colour at maximum and 
decline rate parameter $\Delta m_{15}(B)$ \citep{phi99,alt04}. It gives 
colour excess of $E(B-V)$ = 0.12 for the host galaxy, which is   similar to 
that  obtained using Na\,{\sc i} D lines. The observed discrepancy between Lira's method and 
Phillips/Altavilla's method may be due to difference in the  colour evolution of SN 2012dn and  normal SNe Ia. 
For further analysis we have adopted total reddening 
$E(B-V)_\text{total}$ = 0.18 mag.

\begin{figure}
\resizebox{\hsize}{!}{\includegraphics{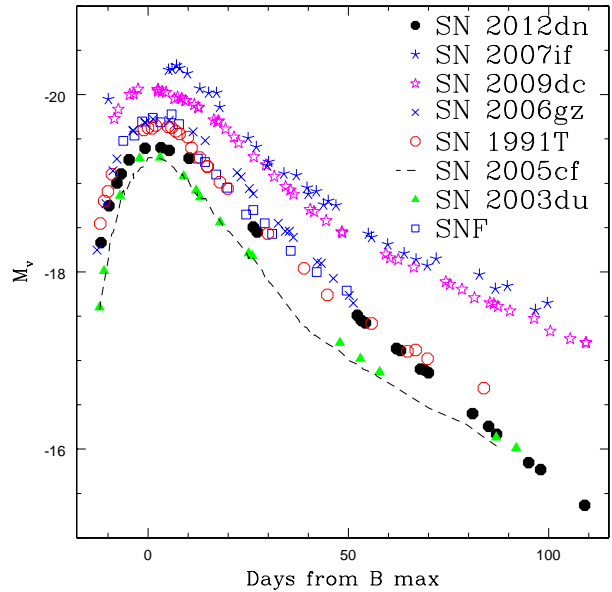}}
\caption[]{The absolute $V$ band light curve of SN 2012dn plotted along with those of SN 2006gz, SN 2007if, SN 2009dc, SN 1991T, SN 2005cf, SN 2003du and SNF 20080723-012. (A colour version of this figure is available in the online journal.)}
\label{fig_absmag}
\end{figure}

Recession velocity of host galaxy ESO 462-16 (PGC 64605), 
corrected for infall of the Local Group towards Virgo Cluster is 
$v_\text{Virgo}$ = 3070 km\,s$^{-1}$ (NED). 
Using $H_0$ = 72 km\,s$^{-1}$\,Mpc$^{-1}$ \citep{fre01} we obtained distance modulus of 
host galaxy of SN 2012dn to be $\mu$ = 33.15 $\pm$ 0.15 mag. Following reddening 
law of \citet{car89} and  $R_V = 3.1$, the peak absolute magnitude of SN 2012dn in 
$U$, $B$, $V$, $R$ and $I$ bands are calculated and given in Table \ref{tab_parameter}. 

The use of $R_V$ as  3.1 has been debated in the recent past. 
There are  studies indicating  value of $R_V$ lower than 3.1 for SNe Ia host galaxies \citep{alt04,rei05,wan06,ama14}. It is  shown by \citet{jha07} and \citet{fol10} that  value of 
$R_V$  lower than 3.1 is preferred  by SNe Ia significantly reddened by dust in their host galaxies.
In  another study \citet{cho11} have also shown that the empirical reddening law, 
derived using optical spectrophotometric data of SNe Ia is compatible with the classical 
extinction law with  $R_V$ = 2.8 $\pm$ 0.3.  
However, for the sake of completeness, the peak absolute magnitude of SN 2012dn is also estimated with 
a lower value of  $R_V$ = 2.3 \citep{wan06}.  With  $R_V$ = 2.3, the absolute magnitude 
of SN 2012dn  will be fainter by 0.151, 0.128, 0.096, 0.072 and 0.046 mag, in $U$, $B$, $V$, $R$ 
and $I$ bands, respectively, which does not change our conclusion. For rest of the analysis we have used $R_V$ as 3.1.

The absolute $V$ band light curves of SN 2012dn and other well studied SNe used for comparison in section \ref{sec:light_curve} are plotted in Fig. \ref{fig_absmag}. 
It is evident that in early phase SN 2012dn is marginally 
brighter  than  normal SNe Ia SN 2003du, SN 2005cf and less luminous
 than SN 1991T and other super-Chandra events. Though in optical bands SN 2012dn  
is marginally bright than the normal type Ia events,  in the UVOT \   $uvw1$ 
and $uvm2$  bands it is $\sim$ one and two magnitudes brighter than normal events, 
respectively \citep{brown14}. During  20--65 d after $B$ 
band maximum, in optical bands SN 2012dn  is  considerably brighter than normal  
SNe Ia and comparable to that of SN 1991T.  
Later on SN 2012dn  declines faster and at $\sim$  100 d its luminosity 
is similar to the normal SNe Ia. 

\begin{figure}
\resizebox{\hsize}{!}{\includegraphics{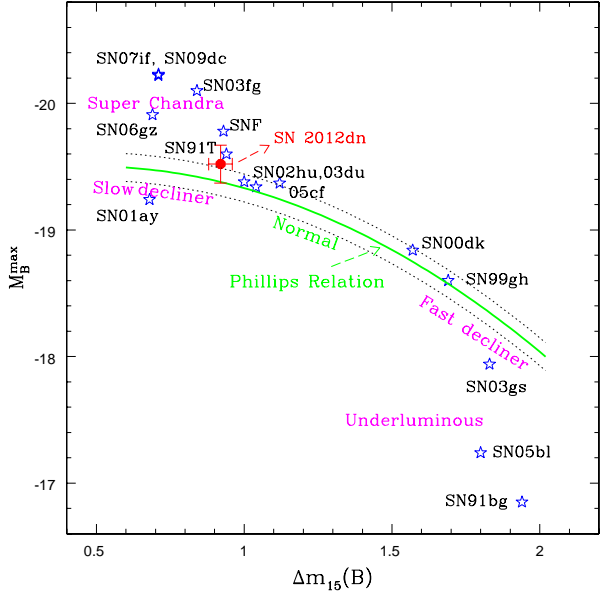}}
\caption[]{Absolute $B$ magnitude ($M_{B}^\text{max}$) is plotted against decline rate parameter $\Delta m_{15}(B)$ for SN 2012dn along with the super-Chandra,  luminous, normal and underluminous SNe Ia. Solid  line represents the luminosity decline rate relation of \citet{phi99}, the 1$\sigma$ uncertainty is represented by dotted lines. (A colour version of this figure is available in the online journal.)}
\label{fig_phillips}
\end{figure}

In Fig. \ref{fig_phillips}, absolute $B$ band magnitude at maximum is plotted against decline rate parameter $\Delta m_{15}(B)$ for
SN 2012dn and other well studied SNe Ia used for comparison in section \ref{sec:light_curve}.   
To have  the observed range of $\Delta m_{15}(B)$ for SNe Ia, some more objects, SN 2001ay \citep{kri11}, SN 2002hu \citep*{sah06}, SN 2003gs \citep{kri09}, SN 2003fg \citep{how06},
SN 2005bl, SN 1999gh, SN 2000dk and SN 1991bg \citep{tau08} are included  in the figure.  For comparison, absolute magnitude corresponding to $R_V$ = 3.1 (as estimated by  authors of the original papers) is taken. For SN 2001ay $R_V$ was taken as 2.4 (Krisciunas et al. 2011), we recalculated  its absolute magnitude  for $R_V$ = 3.1
The luminosity decline rate relation of \citet{phi99} is also  shown in the same figure as solid line. Normal SNe Ia with $\Delta m_{15}(B)$ close to 1.1 follow Phillip's relation \citep{phi99}. 
SNe Ia having  $\Delta m_{15}(B)$ on the extreme ends (overluminous/super-Chandra  and underluminous objects) deviate from  Phillips relation. SN 2012dn  also lies away  from Phillip's relation, although, 
 the deviation is not as significant as is seen in the case of other  super-Chandra objects. 

\subsection{Bolometric light curve}
\begin{figure}
\resizebox{\hsize}{!}{\includegraphics{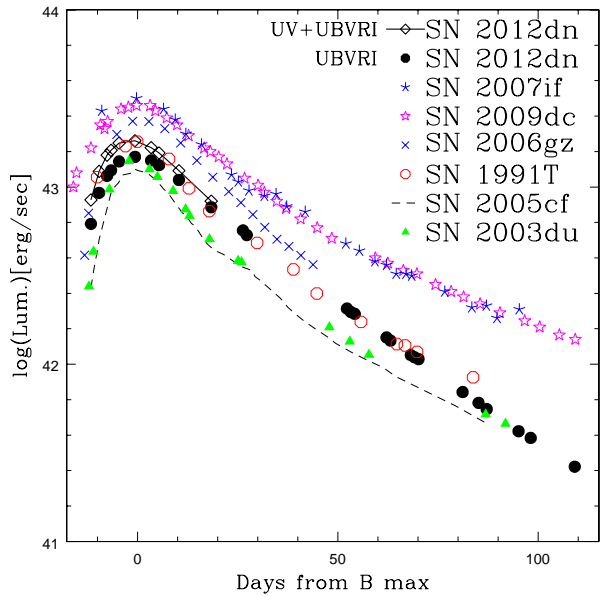}}
\caption[]{Quasi-bolometric light curve  of SN 2012dn compared with those of SN 2006gz, 
SN 2007if, SN 2009dc, SN 1991T, SN 2005cf and SN 2003du. (A colour version of this figure is available in the online journal.)}
\label{fig_bol}
\end{figure}
The observed $U$, $B$, $V$, $R$ and $I$ magnitudes and UVOT $uvw2$, $uvm2$, $uvw1$ magnitudes have been used for deriving the bolometric 
light curve of SN 2012dn.   The optical and UVOT magnitudes are corrected for total 
reddening of $E(B-V)$ = 0.18 mag  using the reddening law of \citet{car89} with $R_V$ = 3.1.  
A distance modulus of $\mu$ = 33.15 $\pm$ 0.15 mag is used. On a  few nights when there is 
no $U$ band observation, data points are obtained by interpolating  the neighbouring 
points. Reddening corrected optical magnitudes were converted to corresponding flux following 
\citet*{bes98}. First we have estimated quasi-bolometric fluxes using the optical bands only, 
 by fitting  spline curve to the $U$, $B$, $V$, $R$ and $I$ fluxes and integrating over the wavelength range 3100 \AA\, to 10600 \AA\, determined by the response  of the optical filters used for the observation. 

The quasi-bolometric light curve of SN 2012dn along with the bolometric light curve of 
SN 2006gz, SN 2007if, SN 2009dc, SN 1991T, SN 2005cf and SN 2003du is plotted in Fig. \ref{fig_bol}. 
The bolometric light curve of SN 1991T, SN 2003du and SN 2005cf are 
computed in a similar fashion as described for SN 2012dn. The bolometric light curve of
SN 2006gz is taken from \citet{hic07}, SN 2007if is taken from \citet{sca10}, and SN 2009dc is taken from \citet{tau11}. From the figure it is evident that the  peak bolometric luminosity of SN 2012dn is greater than those of the normal SNe Ia, comparable to SN 1991T and  less than the super-Chandra SNe Ia 
SN 2006gz, SN 2007if and SN 2009dc. 

In the next step we have included the UVOT fluxes also for 
estimating the quasi-bolometric flux. The extinction in the {\it Swift} UVOT bands have been estimated using the empirical relation and the co-efficients provided by \citet{brown10}. The zero points for
converting the UV magnitudes to fluxes have been taken from \citet{poo08}. The UV-optical quasi-bolometric flux of SN 2012dn was estimated by integrating the monochromatic flux over the wavelength range 1600\AA \ to 10600\AA, and is plotted in Fig. \ref{fig_bol}  with open diamond symbol connected by solid line. 

The peak  quasi bolometric luminosity of SN 2012dn using only optical bands  is  $\log L_\text{bol}^\text{max}$ = 43.17 $\pm$ 0.06 erg s$^{-1}$, while after including the {\it Swift} UV flux the peak quasi-bolometric luminosity  increases by $\sim$ 20\% and  it   goes to  43.25 $\pm$ 0.06 erg s$^{-1}$.  To obtain the $uvoir$ (UV-Optical-IR)  bolometric luminosity,  correction should be made to account for flux in  NIR passbands. \citet{sun96} estimated that contribution from NIR band is at most 10\% at early times. For SN 2005cf, \citet{wan09}  showed  that the NIR contribution declines initially with a minimum of $\sim$ 5\% \citep[see Fig. 23 in][]{wan09} at  $\sim$ 4 d after the $B$ band maximum and later 
it rises linearly reaching to a peak of $\sim$ 20\% at $\sim$ 30 d (during secondary maximum). At nebular phases, the NIR contribution was found to decline gradually reaching below 10\% at $\sim$ 80 d. For SN 2009dc, \citet{tau11} estimated the NIR  contribution of $\sim$ 10\% to the bolometric flux around maximum light. The NIR correction for SN 2007if was less than 5\% around $B$ band maximum \citep{sca10}.  Adding 5\% contributions to account for the missing NIR flux at peak luminosity, we find peak bolometric luminosity of $\log L_\text{bol}^\text{max}$ = 43.27 $\pm$ 0.06 erg s$^{-1}$ for SN 2012dn. 

\subsection{Mass of $^{56}$Ni }
Light curve of SNe Ia is powered by the radioactive decay of $^{56}$Ni synthesized in the explosion. The $^{56}$Ni decays to $^{56}$Co, which subsequently decays to $^{56}$Fe. Mass of $^{56}$Ni can be estimated using Arnett's rule \citep{arn82}, which states that the peak bolometric luminosity of a type Ia SN is proportional to the instantaneous rate of energy release from  radioactive decay chain $^{56}$Ni $\rightarrow$ $^{56}$Co $\rightarrow$ $^{56}$Fe. This can be written in the mathematical form as
\begin{equation*}
\text{Mass of } ^{56}\text{Ni, } \qquad  \text{M}_\text{Ni} = \frac{L_\text{bol}^\text{max}}{\alpha \dot{S}(t_{R})}
\end{equation*}
where $\alpha$ is the ratio of bolometric to radioactive luminosities (near unity) and $\dot{S}(t_{R})$ is the radioactivity luminosity per unit nickel mass  evaluated for the rise time $t_{R}$. From  \citet{nad94}, $\dot{S}(t_{R})$ can be written as
\begin{eqnarray*}
\dot{S}(t_{R}) = \Bigl (6.45~e^{-(t_{R}/8.8d)} + 1.45~e^{-(t_{R}/111.3d)}\Bigr ) \\ 
\times  10^{43} ~~ \text{erg s}^{-1}\text{ M}_\odot^{-1}
\end{eqnarray*}
where 8.8 and 111.3 d are  {\it e}-folding lifetimes ($\tau$) of  $^{56}$Ni and $^{56}$Co, respectively.

SN 2012dn was discovered on 2012 July 08.52 (UT) which is $\sim$ 15.87 d before $B$ band maximum. This provides a lower limit to the rise time. We do not have good constrain on the upper limit as non-detection is reported on 2011 April 01, much before the discovery \citep{boc12}.  The typical SNe Ia have $B$ band rise time of $\sim$ 19 d \citep{con06}. 
For super-Chandra SNe Ia, there is a considerable variation in the  rise time  e.g.  22 d for SN 2003fg \citep{how06}, 18.5 d for SN 2006gz \citep{hic07}, 24.2 d for SN 2007if \citep{sca10} and 23 d for SN 2009dc \citep{sil11}. 
\citet*{gan11} found $B$ band rise time of $\sim$ 18 d for spectroscopically normal SNe Ia and  longer rise time for  slow-declining, luminous SN 1991T-like objects. Their sample of 1991T-like SNe Ia shows rise time distribution near $\sim$ 21 d.
\citet{sca12} also found similar result for luminous objects.
The bolometric light curve of SN 2012dn peaks $\sim$ 1 d before the $B$ band light curve.  
Hence,  we use  rise time of 20 d for SN 2012dn, and find M$_\text{Ni}$ = $\tfrac{0.98}{\alpha}$ M$_\odot$ for peak bolometric luminosity of $\log L_\text{bol}^\text{max}$ = 43.27 erg s$^{-1}$. Arnett's rule uses $\alpha$ = 1 exactly. However, depending on the explosion model the value of $\alpha$ may range from 0.8 to 1.4, with 1.2 as most applicable value \citep{bra92}. Using $\alpha$ = 1.2, for SN 2012dn, we find M$_\text{Ni}$ = 0.82 $\pm$ 0.12 M$_\odot$. The derived mass of $^{56}$Ni is on the higher side of those observed in normal SNe Ia. 
\begin{figure}
\resizebox{\hsize}{!}{\includegraphics{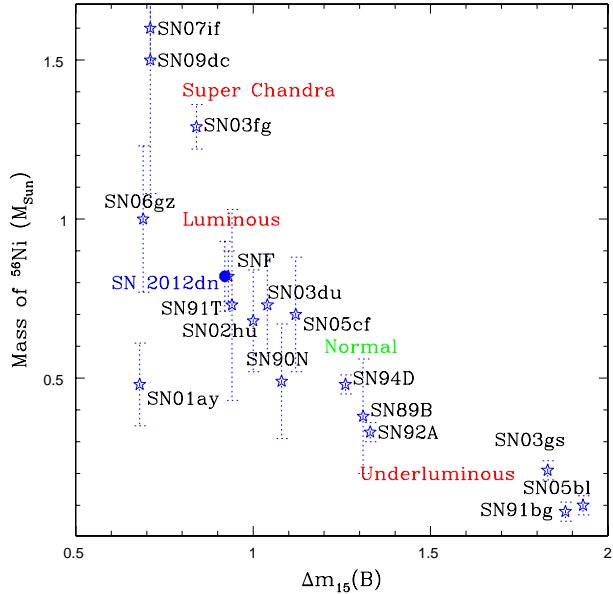}}
\caption[]{Mass of $^{56}$Ni is plotted against decline rate parameter $\Delta m_{15}(B)$ for SN 2012dn along with the super-Chandra,  luminous, normal and underluminous SNe Ia. (A colour version of this figure is available in the online journal.)}
\label{fig_bolNi56}
\end{figure}

In Fig. \ref{fig_bolNi56}, mass of $^{56}$Ni is plotted against $\Delta m_{15}(B)$ for 
SN 2012dn and other well studied SNe Ia used in section \ref{sec:reddening} and Fig. \ref{fig_phillips}. For the sake
of completeness   SN 1989B, SN 1990N, SN 1991T, SN 1991bg,  SN 1992A and 
 SN 1994D were included in this figure, and mass of $^{56}$Ni for these events  were taken from \citet{str06}.
For some of the SNe,  mass of $^{56}$Ni was arrived at by modelling of nebular spectra/light curves 
 as well as by the Arnett's rule. In such cases we  have used the mass of $^{56}$Ni estimated using  Arnett's rule.  
Whenever required the reported  mass of $^{56}$Ni was scaled for  $\alpha$ = 1.2.
It is found that SN 2012dn lies in the luminous SNe Ia group, consistent with its other 
observed properties.

\section{Spectral evolution}
\label{sec:spec_evol}
Medium resolution spectra of SN 2012dn observed during  $-$11.6 d  to +98.2 d with respect to  
$B$ band maximum are presented in Fig. \ref{fig_spec1}, \ref{fig_spec2}, \ref{fig_spec3} and   \ref{fig_spec4}. 
The spectra have been  corrected for redshift ({\it z }= 0.01) and total reddening of $E(B-V)_\text{tot}$ = 0.18 as described in sections \ref{sec:spec_observation} and \ref{sec:reddening}. 
\subsection{Pre-maximum}
\label{sec:premax_spec}
\begin{figure}
\resizebox{\hsize}{!}{\includegraphics{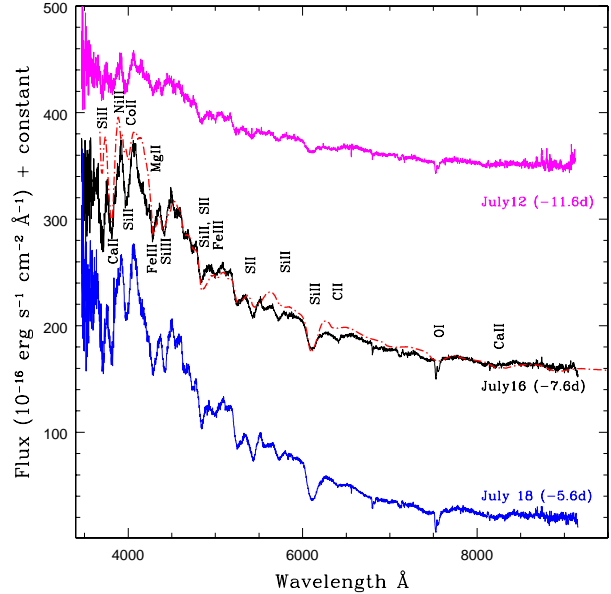}}
\caption[]{Pre-maximum spectral evolution of SN 2012dn during $-$11.6 d to $-$5.6 d with respect to $B$ band maximum. Synthetic spectrum generated using the {\sc syn++} code is also plotted (dashed line) with the $-$7.6 d spectrum. (A colour version of this figure is available in the online journal.)}
\label{fig_spec1}
\end{figure} 
The pre-maximum spectra of SN 2012dn obtained at $-$11.6, $-$7.6 and $-$5.6 d with respect to 
 $B$ band maximum 
are plotted in Fig. \ref{fig_spec1}. The first spectrum obtained on 
 $-$11.6 d, has blue continuum with relatively narrow absorption lines.
The characteristic absorption features of SNe Ia due to Si\,{\sc ii} $\lambda$6355, 
$\lambda$5972,  $\lambda$4130 and $\lambda$3858,  S\,{\sc ii} lines at $\lambda$5454, 
$\lambda$5640 and narrow  
Ca\,{\sc ii} H\&K lines are clearly seen. The blueshift of the Si\,{\sc ii} $\lambda$6355 line is 
$\sim$ 11900 km\,s$^{-1}$. There is a broad absorption redward of Si\,{\sc ii} $\lambda$6355 line, 
which is associated with the  C\,{\sc ii} $\lambda$6580 line \citep{par12}. 
This line is blueshifted by $\sim$ 9500 km\,s$^{-1}$. 
Other lines due to C\,{\sc ii} at $\lambda$7234 and  $\lambda$4267 are marginally detected 
in this spectrum. 
The other two spectra at  $-$7.6 and $-$5.6 d  are similar to 
that of  $-$11.6 d. The  absorption  lines have become stronger. The C\,{\sc ii} 
$\lambda$6580 line is detected in all the pre-maximum spectra
with reducing  strength while  the other C\,{\sc ii} lines are not seen in $-$7.6 
and $-$5.6 d spectra. The Ca\,{\sc ii} NIR triplet is becoming stronger. The 
Si\,{\sc iii} $\lambda$4560 line is very strong and its strength is comparable 
to Mg\,{\sc ii} line in the pre-maximum spectra. 

Pre-maximum spectrum at $-$7.6 d  is fit  with the synthetic spectrum generated using the 
{\sc syn++}{\footnote{\url{https://c3.lbl.gov/es/}}} code and plotted in Fig. \ref{fig_spec1}. 
The {\sc syn++} code is a rewrite and enhanced version of original parametrized supernova 
synthetic spectrum code {\sc synow} \citep{fis00}, with more complete atomic data files 
\citep*{tho11}. The {\sc synow} code is based on basic assumption of spherical symmetry 
where velocity of ejecta is  proportional to radius ($v \propto r$). Line formation is 
treated using the Sobolev approximation \citep{sob57,jef90} and occurs purely due to 
resonant scattering outside a sharp photosphere that emits blackbody continuum at a 
given temperature. Profile of a reference line of an ion is determined by optical 
depth which is a function of velocity. Relative strengths of  other lines of the same 
species are calculated  assuming Boltzmann statistics  at an excitation temperature. 

The $-$7.6 d spectrum of SN 2012dn matches well with the synthetic spectrum  
having  photospheric velocity of 11800 km\,s$^{-1}$ and blackbody temperature of 26000 K. 
The synthetic spectrum with lower temperature ($\sim$ 12000 K),  used for normal 
SNe  Ia  \citep{branch05} does not fit the  bluer part (below 5500\AA) of the spectrum. 
An excess emission is noticed in the bluer part of the spectrum, consistent with the very blue
$(uvw1-v)$ and $(U-B)$ colours of SN 2012dn  in the pre-maximum phase.  The synthetic spectrum includes ions of O\,{\sc i}, Mg\,{\sc ii}, Si\,{\sc ii}, 
Si\,{\sc iii}, 
S\,{\sc ii}, Ca\,{\sc ii}, Fe\,{\sc ii}, Fe\,{\sc iii}, Co\,{\sc ii} and Ni\,{\sc ii}. 
Optical depths of all the lines were set to decrease exponentially with velocity, keeping 
e-folding velocity at 1000 km\,s$^{-1}$.
An excitation temperature of 10000 K was used for all ions except C\,{\sc ii}, for 
which it was taken as 17000 K to match with the relative line strength in the observed spectrum.   This is similar to the case of SN 2009dc and SN 2003fg wherein   higher  
excitation temperatures of 20000 K and 35000 K  were used \citep{sil11,how06}. 
The lines identified with the help of  synthetic spectrum are marked in Fig. \ref{fig_spec1}. 
Inclusion of Co\,{\sc ii} and Ni\,{\sc ii} improves the fit at Si\,{\sc ii} $\lambda$4130.

The  feature  at emission part of W shaped S\,{\sc ii} and blueward to 
Si\,{\sc ii} $\lambda$5972 in the observed spectrum could not be reproduced in the  
synthetic spectrum. 
A similar mismatch was also noted by \citet{tau11} for SN 2009dc at $\sim$ +7 d.
A  blend of Na\,{\sc i} D with Si\,{\sc ii} $\lambda$5972 was suggested to be responsible 
for this feature  \citep{tau11}.  To associate this feature with Na\,{\sc i} D in SN 2012dn, 
relatively high ejecta velocity is required. 
Si\,{\sc iii} with significantly lower photospheric velocity could also give rise to 
this feature. However, with the lower photospheric velocity,  lines due to other ions 
become much  stronger than seen in the  observed spectrum.

\begin{figure}
\resizebox{\hsize}{!}{\includegraphics{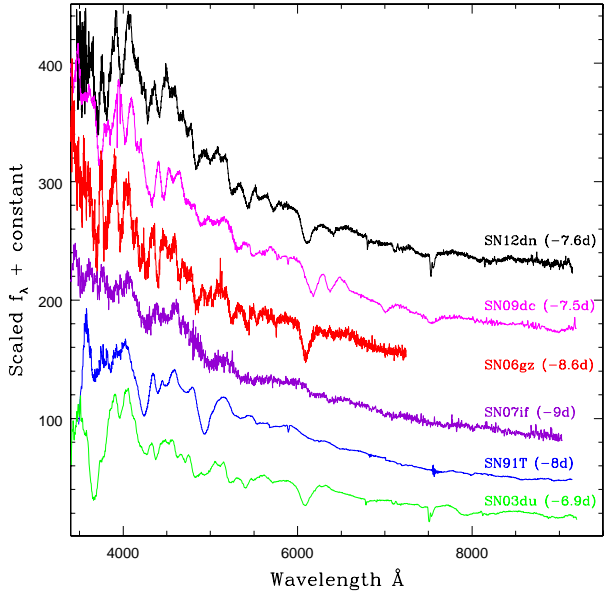}}
\caption[]{Comparison of pre-maximum spectrum of SN 2012dn at $-$7.6 d with those of SN 2009dc, SN 2006gz, SN 2007if, SN 1991T and SN 2003du at similar epoch. (A colour version of this figure is available in the online journal.)}
\label{fig_spec_comp1}
\end{figure} 

The $-$7.6 d spectrum of SN 2012dn is plotted in Fig. \ref{fig_spec_comp1} with spectra of 
SN 2009dc \citep{tau11}, SN 2006gz \citep{hic07}, SN 2007if \citep{sca10}, 
SN 1991T \citep{jef92,sch94} and SN 2003du \citep{anu05} taken around similar epoch.  
The spectra of SN 2009dc (used in this figure only) and SN 2007if are taken  from 
Weizmann Interactive Supernova Data Repository 
(WISeREP{\footnote{\url{http://www.weizmann.ac.il/astrophysics/wiserep/}}}).
Spectra of SN 2006gz have been obtained from CfA Supernova Data 
Archive{\footnote{\url{http://www.cfa.harvard.edu/supernova/SNarchive.html}}} 
while  spectra of SN 1991T are obtained from CfA and 
SUSPECT{\footnote{\url{http://www.nhn.ou.edu/~suspect/}}} Supernova Spectrum Archive. 
Spectra of SN 2009dc except at  $\sim -$7 d are from our own 
observations with the HCT. A close resemblance between the spectrum of SN 2012dn 
and SN 2006gz is very clear. Spectra of both the objects are characterized by 
well developed Si\,{\sc ii} $\lambda$6355, strong Si\,{\sc iii} $\lambda$4560, 
weak C\,{\sc ii} $\lambda$6580 and very narrow 
lines due to other IMEs.  Spectrum of  SN 2003du at $\sim -$7 d 
shows lines due to the species seen in   the spectra of SN 2012dn and SN 2006gz,
however, the relative line strength and the line widths are different from those in SN 2012dn.
In SN 2003du the lines are broader and  blended more. Though,  the decline rate parameter of 
SN 2012dn and SN 1991T are similar, they show  considerable difference in the pre-maximum
spectral evolution.  Spectrum of  SN 1991T at  $\sim -$8 d doesnot show the 
Si\,{\sc ii} $\lambda$6355 line, instead it shows  prominent Fe\,{\sc iii} lines. 
SN 2007if seems to have featureless spectrum similar to SN 1991T. The  Fe\,{\sc iii} 
lines are even weaker in the spectrum of SN 2007if as compared to SN 1991T.  

\subsection{Maximum to early nebular phase}
\begin{figure}
\resizebox{\hsize}{!}{\includegraphics{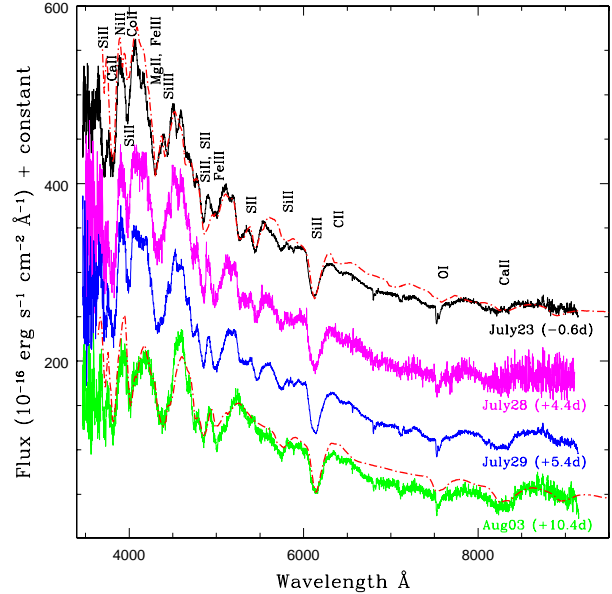}}
\caption[]{Spectral evolution of SN 2012dn during close to $B$ band maximum to +10 d. Synthetic spectra  generated using the {\sc syn++} code are also plotted (dashed line) with  $-$0.6 d and +10.4 d spectra. (A colour version of this figure is available in the online journal.)}
\label{fig_spec2}
\end{figure} 

\begin{figure}
\resizebox{\hsize}{!}{\includegraphics{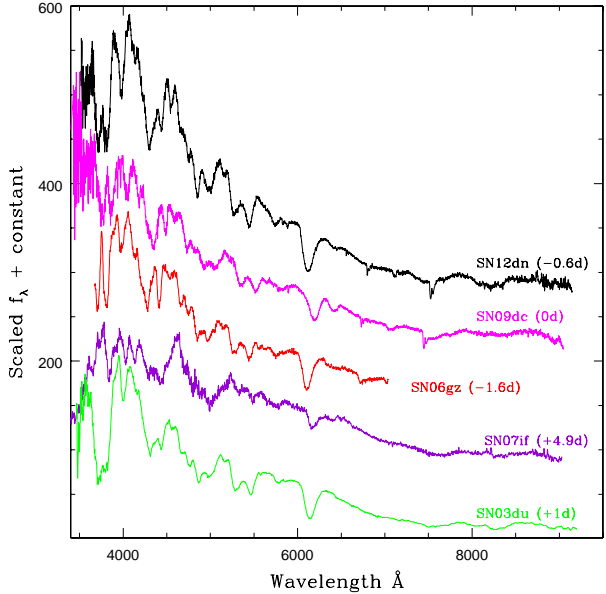}}
\caption[]{Spectrum of SN 2012dn close to $B$ band maximum is plotted with spectra of SN 2009dc, SN 2006gz,
SN 2007if and SN 2003du at similar epoch. (A colour version of this figure is available in the online journal.)}
\label{fig_spec_comp2}
\end{figure} 

\begin{figure}
\resizebox{\hsize}{!}{\includegraphics{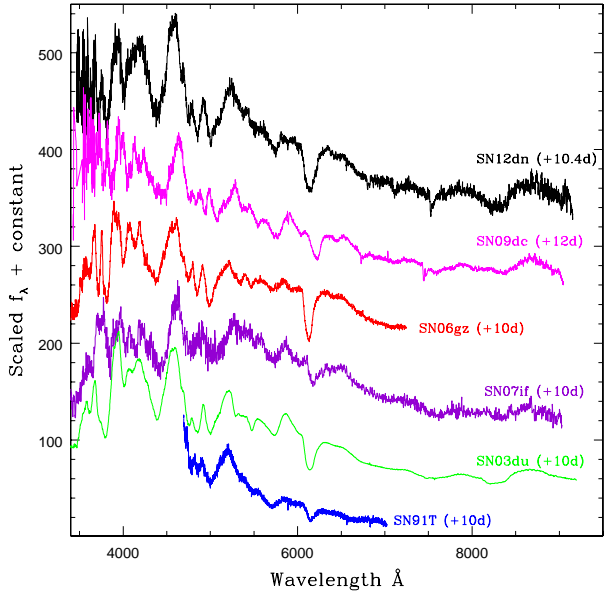}}
\caption[]{Spectrum of SN 2012dn at +10.4 d is compared with those of SN 2009dc, SN 2006gz, SN 2007if, SN 2003du and SN 1991T at similar epoch. (A colour version of this figure is available in the online journal.)}
\label{fig_spec_comp3}
\end{figure} 

The spectral evolution during early post-maximum phase (close to maximum  to 10 d after maximum)
is displayed  in Fig.  \ref{fig_spec2}. As compared to the $-$5.6 d, the spectrum close to maximum 
has redder continuum, which becomes increasingly redder at  later epochs. 
The narrow Si\,{\sc ii} $\lambda$4130 and  Ca\,{\sc ii}  H\&K lines are well resolved and 
a  small notch with decreasing strength at the position 
of C\,{\sc ii} $\lambda$6580 is seen till +10.4 d. The strong Si\,{\sc iii} $\lambda$4560 line 
in the pre-maximum
spectra starts blending with the Mg\,{\sc ii} line and disappears $\sim$ 5 d after $B$ band maximum. 
The S\,{\sc ii} lines  weaken and disappear in +10.4 d spectrum.  The  blend 
of  Na\,{\sc i} with   Si\,{\sc ii} at $\sim$ 5800 \AA \ and Ca\,{\sc ii} NIR triplet start 
becoming stronger.
 
In Fig. \ref{fig_spec2} spectra of SN 2012dn close to $B$ band maximum and at +10.4 d, are
 compared with the synthetic spectra obtained with {\sc syn++}. As noted with the 
spectrum at $-$7.6 d, the synthetic spectrum with the blackbody temperature ($\sim$ 13000 K) 
used for normal SNe Ia does not fit the bluer part of the observed spectrum at $B$ maximum.
 A synthetic spectrum with  
 photospheric velocity of 11000 km\,s$^{-1}$ and blackbody temperature of 20000 K reproduces 
  the shape  and features seen in the observed spectrum reasonably well.  
The ions included and their lines are marked in Fig. \ref{fig_spec2}. 
The observed spectrum at +10.4 d is found to match   well with the synthetic spectrum
 having  photospheric velocity of 10500 km\,s$^{-1}$ and blackbody temperature 
of 12000 K, similar to normal SNe Ia at similar  epoch. An excitation temperature 
of 7000 K was used for all the ions. The synthetic spectrum at +10.4 d includes 
contributions from O\,{\sc i}, Na\,{\sc i}, 
Mg\,{\sc ii}, Si\,{\sc ii},  Ca\,{\sc ii}, Fe\,{\sc ii} and Co\,{\sc ii} ions. 

The spectra  of SN 2012dn  and  other well studied SNe around maximum are compared  in Fig. \ref{fig_spec_comp2}. 
Spectra of SN 2012dn, SN 2006gz appear very similar with narrow lines of IMEs. However, 
they differ 
considerably from the spectrum of SN 2009dc.  The C\,{\sc ii} $\lambda$6580 line is prominent
in SN 2009dc, whereas it is marginally seen in the spectra of SN 2012dn and SN 2006gz.
The lines  due to Si\,{\sc ii} $\lambda$6355, Ca\,{\sc ii} H\&K  and 
Ca\,{\sc ii} NIR triplet are weak in SN 2009dc as compared to SN 2012dn and SN 2006gz. 
Lines in the spectrum of SN 2003du continues to be broader than those in SN 2012dn. 
In SN 2007if  the Si\,{\sc ii} and other spectral  features of IMEs start  developing 
around  +5 d.  In Fig. \ref{fig_spec_comp3}, spectrum of SN 2012dn at +10.4 d is 
plotted with those of other SNe at similar epoch.  The  spectral similarity  of SN 2012dn, 
 SN 2009dc and  SN 2006gz is apparent, they all show narrow lines. As compared to SN 2009dc,  
Si\,{\sc ii} line is  deeper in SN 2012dn and SN 2006gz.  Spectral features of SN 2007if 
are shallower than all the SNe in comparison.   

\begin{figure}
\resizebox{\hsize}{!}{\includegraphics{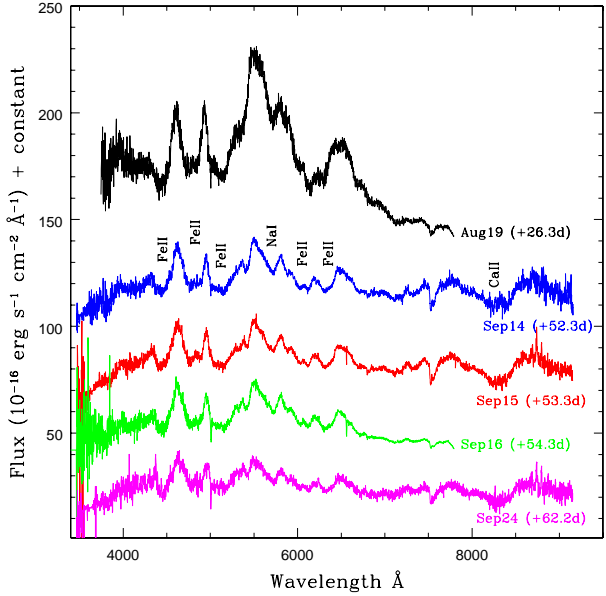}}
\caption[]{Spectral evolution of SN 2012dn during +26.3 d to +70.2 d with respect to $B$ band maximum. (A colour version of this figure is available in the online journal.)}
\label{fig_spec3}
\end{figure}

\begin{figure}
\resizebox{\hsize}{!}{\includegraphics{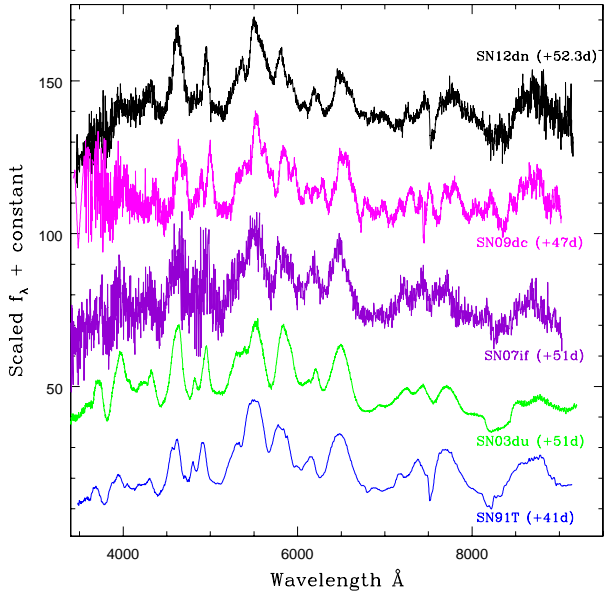}}
\caption[]{Comparison of spectrum of SN 2012dn at +52.3 d with those of SN 2009dc, SN 2007if, SN 2003du and SN 1991T at similar epoch. (A colour version of this figure is available in the online journal.)}
\label{fig_spec_comp4}
\end{figure}

The spectral evolution of SN 2012dn during $\sim$ 1  to 2 months after $B$ band maximum 
is displayed in Fig. \ref{fig_spec3}.  
The Si\,{\sc ii} $\lambda$6355 feature is  replaced  by   developing Fe\,{\sc ii} lines in 
spectrum of SN 2012dn obtained at +26.3 d.  The  
Na\,{\sc i} has also strengthened by this time. The spectra around  50 d after $B$ 
band maximum  lack in Si\,{\sc ii},  it is now dominated by Fe\,{\sc ii}. The other 
strong features are due to Na\,{\sc i} and Ca\,{\sc ii} NIR triplet.
Spectrum of SN 2012dn at +52.3 d is compared with  spectra of other SNe at similar 
epoch in Fig. \ref{fig_spec_comp4}. Similar to SN 2009dc, the spectral features of 
SN 2012dn are relatively narrow than those of SN 2003du and SN 1991T. 
As compared to SN 2003du, the Ca\, {\sc ii} H\&K and Ca\, {\sc ii} NIR triplet  
are weaker in SN 2012dn. 
  
\begin{figure}
\resizebox{\hsize}{!}{\includegraphics{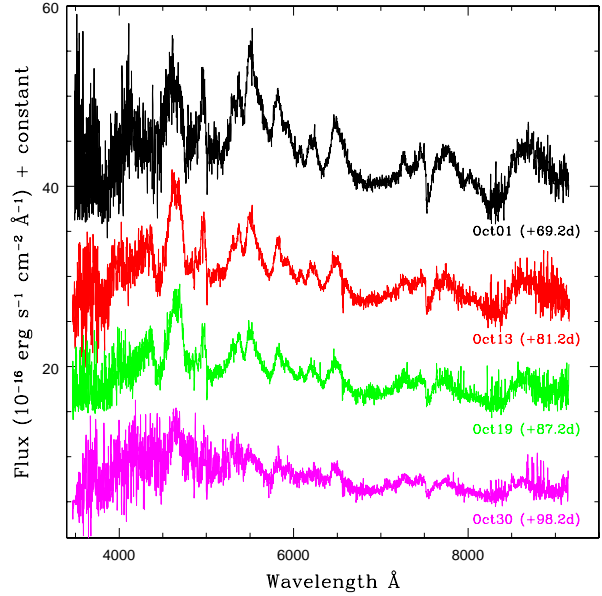}}
\caption[]{Spectral evolution of SN 2012dn during +69.2 d to +98.2 d with respect to $B$ band maximum. (A colour version of this figure is available in the online journal.)}
\label{fig_spec4}
\end{figure}

\begin{figure}
\resizebox{\hsize}{!}{\includegraphics{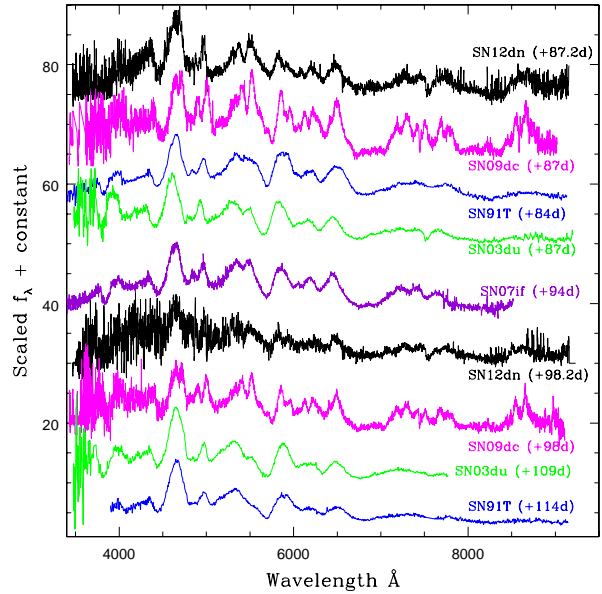}}
\caption[]{Spectra of SN 2012dn at +87.2 d and +98.2 d have been compared with those of SN 2009dc, SN 2007if, SN 1991T and SN 2003du at similar epochs. (A colour version of this figure is available in the online journal.)}
\label{fig_spec_comp5}
\end{figure}

The spectral evolution of SN 2012dn in early nebular phase during +69.2 to +98.2 d is shown in Fig. \ref{fig_spec4}. There is no appreciable change in the shape of spectra. Spectra are dominated by forbidden  emission lines  [Fe\,{\sc ii}], [Fe\,{\sc iii}] and [Co\,{\sc iii}]. Spectra of SN 2012dn at +87.2 and +98.2 d are compared with those of SN 2009dc, SN 2007if, SN 1991T and SN 2003du at similar epoch in Fig. \ref{fig_spec_comp5}. With narrow lines, spectrum of SN 2012dn at +87.2 d shows similarity with that of SN 2009dc, whereas other SNe in comparison show broad features. Further, in the red region (beyond 
7000 \AA)   SN 2012dn shows  emission features which  are absent in the spectra of SN 1991T and  SN 2003du. 
The emission features in the region $\sim$ 7300--7800 \AA\,  and  Ca\,{\sc ii} NIR triplet emission are stronger in 
SN 2009dc. The emission features around  7500 \AA\, are attributed  to [Fe\,{\sc ii}]/[Ni\,{\sc ii}] and [Ca\,{\sc ii}]
in SN 2009dc \citep{tau13}. These emission lines are seen in the spectrum of SN 2012dn also  but with lower strength, similar
to SN 2007if. The emission feature around 5500 \AA\ generally assigned  to [Fe\,{\sc iii}] and [Fe\,{\sc ii}] lines 
 are well separated in the spectra of SN 2012dn and SN 2009dc,  it is marginally separated in the spectra of SN 2003du 
and SN 2007if, whereas in SN 1991T a flat top blend is seen. 
 
Nebular spectra (t $>$ 300 d) of super-Chandra SNe Ia SN 2006gz, SN 2007if and SN 2009dc have been discussed by \citet{mae09} and \citet{tau13}. The  strong [Fe III] emission around 4700 \AA\,  seen in the nebular  spectra of  normal and SN 1991T-like objects was found to be very weak or absent in the spectra of super-Chandra objects. In the late nebular phase ($\sim$ 300 d),  the flux ratio [Fe III] / [Fe II] of  emission lines near 4700 and 5200 \AA\, was found   in the range 1.0--1.1 for super-Chandra SNe, while  for  normal and SN 1991T-like objects this ratio ranges between 1.3--1.9. Suppression of [Fe III] emission lines was also apparent  in the spectrum of SN 2009dc taken  at +165.1 d \citep{tau11}. Our last spectrum of SN 2012dn  at +98.2 d is too noisy below 5000 \AA\,   to estimate flux ratios.

\subsection{Velocity evolution of the supernova ejecta}
\begin{figure}
\resizebox{\hsize}{!}{\includegraphics{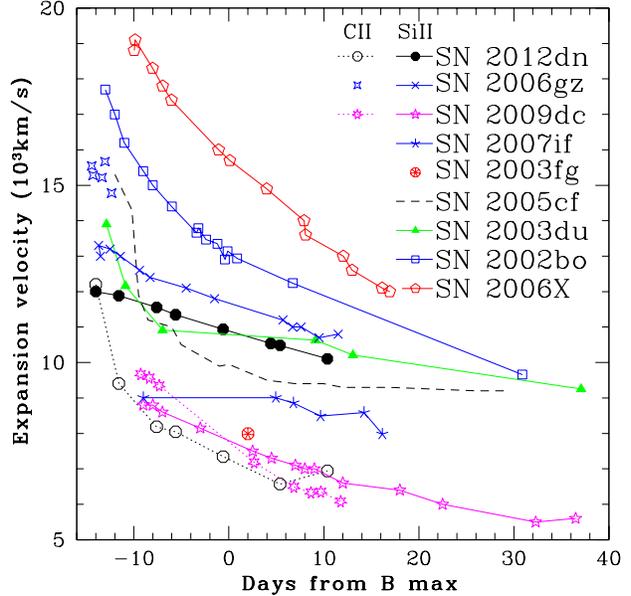}}
\caption[]{Velocity evolution of  Si\,{\sc ii} 6355 \AA\, absorption line for SN 2012dn is compared with those of SN 2006gz, SN 2007if, SN 2009dc, SN 2003fg, SN 2005cf, SN 2003du, SN 2006X and SN 2002bo. Velocity evolution derived using C\,{\sc ii} 6580 \AA\, line is also plotted for SN 2012dn, SN 2006gz and SN 2009dc. (A colour version of this figure is available in the online journal.)}
\label{fig_vel}
\end{figure}

The expansion velocity of SN 2012dn  measured using 
 minimum of the Si\,{\sc ii} $\lambda$6355 absorption line at different epochs, is plotted in Fig. \ref{fig_vel}. The first data point in the plot is from the early velocity measurement  by \citet{par12}.  
Velocity evolution of  other well studied SNe Ia  have also been plotted for 
 comparison in the same figure. During  the pre-maximum phase,  till $\sim$ one week before $B$ band maximum,  the Si\,{\sc ii} velocity of normal SNe Ia shows a rapid decline, which is 
not seen in case of SN 2012dn and other super-Chandra SNe Ia. The  Si\,{\sc ii}
velocity of SN 2012dn declines linearly.  Though the Si\,{\sc ii}  line velocity evolution of  SN 2012dn is similar to other super-Chandra SNe Ia, the measured velocities are significantly different.   The Si\,{\sc ii}   line velocity  of SN 2012dn is higher than that of super-Chandra SNe Ia SN 2009dc, SN 2007if,  SN 2003fg and marginally lower than SN 2006gz. 

At $-$11.6 d the Si\,{\sc ii} velocity for SN 2012dn is $\sim$ 11900 km\,s$^{-1}$ which reduced to  $\sim$ 10900 km\,s$^{-1}$  at $B$  band maximum. Around maximum SN 2006gz has Si\,{\sc ii} velocity $\sim$ 11500 km\,s$^{-1}$, similar to that of a typical type  Ia supernova. The Si\,{\sc ii} velocity  of SN 2007if remained $\sim$ 9000 km\,s$^{-1}$ from pre-maximum phase 
to $\sim$ one week after maximum, exhibiting a plateau-like feature in velocity evolution.  
A small plateau phase during $-$14 to $-$13 d was observed in SN 2006gz \citep{hic07}. 
We donot see any plateau-like feature in Si\,{\sc ii} velocity evolution of SN 2012dn, it declines linearly with a slope similar to SN 2006gz. 
 Slow velocity evolution with  plateau like feature and presence of unburned materials observed in super-Chandra SNe Ia
have been interpreted as a result of envelope/shell or clumpy structure in their ejecta \citep{sca10}. 

The C\,{\sc ii} $\lambda$6580 line velocity of SN 2012dn along with those of SN 2006gz and SN 2009dc is also  
 plotted in the same figure. The first measurement of C\,{\sc ii} line velocity is taken 
from \citet{par12}, wherein they have measured a velocity of 
$\sim$ 12200  km\,s$^{-1}$, marginally higher than the velocity of Si\,{\sc ii} line ($\sim$ 12000 km\,s$^{-1}$). In SN 2006gz, C\,{\sc ii} line was detectable  only during the pre-maximum phase with 
expansion velocity always higher than that of Si\,{\sc ii} line. The C\,{\sc ii} velocity in SN 2009dc was
$\sim$ 800 km\,s$^{-1}$ more than Si\,{\sc ii} velocity around 9 d before $B$ maximum.  Around
maximum light C\,{\sc ii} and Si\,{\sc ii} velocities were comparable and afterwards  C\,{\sc ii} velocity
was lower than Si\,{\sc ii} velocity.  In SN 2012dn C\,{\sc ii} velocity evolution appears peculiar as compared 
to SN 2006gz and SN 2009dc. Except at $-$13 d (when C\,{\sc ii} velocity is marginally higher than 
Si\,{\sc ii} velocity), C\,{\sc ii} velocity is always lower than Si\,{\sc ii} velocity. The C\,{\sc ii} velocity decreases
rapidly from $\sim$12000 km\,s$^{-1}$ to  $\sim$9500 km\,s$^{-1}$ in just 2 days. Afterwards
the difference in velocity estimated using Si\,{\sc ii} and C\,{\sc ii} is $>$ 3000 km\,s$^{-1}$. 

\citet{fol12} have shown that for normal SNe Ia  the velocity evolution of C\,{\sc ii}  is parallel to  that of Si\,{\sc ii} with an offset of   $\sim$ +1000 km\,s$^{-1}$. In another study of  carbon features in SNe Ia, the ratio of Doppler velocities {\it v\,}(C\,{\sc ii} $\lambda$6580)/{\it v\,}(Si\,{\sc ii} $\lambda$6355) has been explored  and for majority of SNe Ia
it is found to be close to 1  \citep{par11}. For SN 2006bt and SN 2008ha, this ratio is
$\sim$ 0.5 suggesting  that the carbon clumps are ejected away from  the line of sight of the observer. \citet{par11} have shown that any observed discrepancy between $v_\text{phot}$ and $v_\text{C\,{\sc ii}}$ could be explained if the actual velocity of C\,{\sc ii} is same as $v_\text{phot}$ but instead of forming  in a shell at the observed velocity, the carbon is  in a clump at $v_\text{phot}$ and offset by an angle from the line of sight.
 It is possible that the observed discrepancy in the Si\,{\sc ii}  and C\,{\sc ii} $\lambda$6580
 velocity in SN 2012dn may be due to clumping and projection effect of carbon rich material, however,
 detailed modelling is required to confirm it.  

\section{Discussion and summary}
\label{sec:summary}
The properties exhibited by  SN 2012dn makes it an interesting object. 
SN 2012dn is a  slow-declining  ($\Delta m_{15}(B)_\text{true}$ = 0.92) and marginally luminous type Ia supernova. The peak $B$ and $V$ band absolute magnitudes are $M_{B}^\text{max}$ = $-$19.52 and $M_{V}^\text{max}$ = $-$19.42, respectively. 
The light curve evolution of SN 2012dn is different from normal SNe Ia in the following way: (i) the light curve in  $I$ band peaks after maximum in $B$ band, which is  opposite to the observed trend for normal type Ia events, (ii) the strength of the primary and secondary maximum is almost similar in SN 2012dn, whereas, in normal SNe Ia the secondary maximum in $I$ is found to be $\sim$ 0.5 mag fainter than the primary maximum and (iii) the late phase decline of SN 2012dn  in all the bands is faster.  Though, it is marginally luminous  in the optical bands, it is $\sim$ one magnitude and two magnitudes brighter than normal SNe Ia  in UVOT $uvw1$ and $uvm2$ bands, respectively. It  shows very blue  $(uvw1-v)$ and $(U-B)$ colours. The contribution of UV bands to the bolometric flux is quite high,  it evolves from $\sim$ 37\% (at $-$12 days)  to $\sim$ 8\% (+18 days). At  $B$ maximum $\sim$ 20\% of the bolometric flux is emitted in UV bands, similar to SN 2009dc \citep{sil11}.   The peak bolometric luminosity of $\log L_\text{bol}^\text{max}$ = 43.27 erg s$^{-1}$ indicates that $\sim$ 0.82 M$_\odot$  of  $^{56}$Ni was synthesized in the explosion. The mass of $^{56}$Ni for SN 2012dn is on the higher side of $^{56}$Ni mass distribution for normal type Ia events.  

SN 2012dn shows  spectral features characteristic  of normal  SNe Ia, but  with relatively narrow absorption lines. Presence of unburned materials in the ejecta is evident from the C\,{\sc ii} 6580 \AA\, absorption feature seen in the pre-maximum spectra. During  the pre-maximum  and close to maximum phase, to reproduce observed shape of the  spectra, the synthetic spectrum code {\sc syn++}  needs significantly higher blackbody temperature,  than those required for normal type Ia events. However, during  the post-maximum phase, synthetic spectrum with blackbody temperature similar to normal type Ia fits the observed spectrum well.  Photospheric velocity of SN 2012dn, inferred from  Si\,{\sc ii} 6355 \AA\, absorption line,  is marginally greater than those of the super-Chandra SNe: SN 2003fg, SN 2007if, SN 2009dc and lower than SN 2006gz and normal type Ia events. Similar to super-Chandra events, SN 2012dn has slow velocity evolution which is almost parallel to that of SN 2006gz.  

Normal SNe Ia are expected to produce only weak emission in the far-UV, but interaction of the supernova ejecta with an extended progenitor such as red-giant star can produce excess of UV photons \citep{kas10}. Based on the smoothness of the UV light curves and their qualitative similarity to the optical light curves, the higher luminosity of SN 2012dn  is associated with  photospheric origin \citep{brown14}.   Further, the presence of  stronger features in MUV (below 2700 \AA) and optical spectra also supports the photospheric origin for the excess emission, as flux from a hot shock would be relatively smooth and would dilute the photospheric features \citep*{brown14, hamuy03}. Hence, it is suggested that a higher temperature and lower opacity may contribute to the UV excess rather than a hot, smooth blackbody from shock interaction \citep{brown14}.  However, the possibility of a structured spectrum with emission and absorption arising  due to reprocessing of  shock emission or originating from a different composition, is not ruled out. They have emphasized the need for higher quality UV spectra at the earliest possible epochs to probe the mechanism responsible for the excess UV emission. The  {\sc syn++} fit of the pre-maximum spectra with very high blackbody temperature also supports the photospheric origin for excess luminosity in the early phase. 

The light generation by significant interaction of the ejecta with a CSM can also result in increased luminosity of super-Chandra SNe \citep*{tau11,hach12}.  Usually  presence of CSM around  SNe Ia  is inferred through either temporal variability \citep{pat07} or statistical analysis  of the velocity of narrow absorption features \citep{stern11}.  
Supernovae with relatively narrow hydrogen emission lines that are linked to SNe Ia are labelled as ``Ia-CSM'' objects.  \citet{sil13} have investigated the observable signatures of SNe Ia-CSM and argued that at least some SNe Ia arise from the SD channel, since  a hydrogen rich CSM is most likely the result of SD scenario. These are the extreme cases of interaction of supernova ejecta with the CSM.  The absence of  any variable narrow absorption lines and  narrow hydrogen  emission lines in our medium resolution spectra indicates  that appreciable amount of  CSM may not be present around  SN 2012dn to give rise to  higher luminosity.  

The decline in the magnitude in 50 days after $B$ maximum listed in the Table \ref{tab_decline} shows that SN 2012dn has slower decline than normal SNe Ia  and SN 1991T and  faster than SN 2009dc in all the bands (in $B$ band SN 1991T and SN 2012dn have comparable decline rates). \citet{sca14} have demonstrated that despite having significantly different  peak magnitudes, shape of the   post-maximum light curve in $g$ band of supernova LSQ12gdj, SN 2007if and  SN 1991T-like SN 2005M are similar and different from the Ia-CSM objects \citep[refer Fig. 10 of][]{sca14}. However, presence of blueshifted Na\,{\sc i}D and  Ca\,{\sc ii} H\&K absorption in the high resolution spectrum of  supernova LSQ12gdj led \citet{sca14} to postulate that some CSM may be present around LSQ12gdj while  its contribution to the luminosity is negligible. \citet{hach12} have discussed the possibility of the CSM consists of C, O or heavier elements, sufficiently  close to the progenitor. The tamped detonation,  in which a white dwarf surrounded by an extended envelope   explodes \citep*[and references therein]{kho93}, presents a possible progenitor system with CSM. It produces luminous events with longer rise time, which appear much like normal type Ia supernova after maximum. The extended envelope may come  from an accreted binary companion.   The presence of  C feature in the spectra of SN 2012dn till $\sim$ 10 days after maximum indicates  possibility of explosion in a C rich environment. The presence of C in the CSM would  also provide conducive environment for dust formation, which may explain the observed   steepening in the  light curve  $\sim$ 50 days after maximum.  A detailed model is required to check these possibilities.       

\section*{Acknowledgment}
We acknowledge insightful comments from the anonymous referee that helped in improving the paper.
NKC is thankful to Indian National Science Academy (INSA), New Delhi for giving an opportunity to work under INSA Visiting Fellowship Scheme (Sanction no. SP/VF-3/2013-14/353) at Indian Institute of Astrophysics (IIA), Bangalore. NKC would like to thank the Director and Dean of IIA for local hospitality and facilities provided. 
We are thankful to the staff at IIA for their assistance during the observations and to all the observers of the 2-m HCT (IAO-IIA), who kindly provided part of their observing time for supernova observations. This work has made use of public data in the Swift data archive and the NASA/IPAC Extragalactic Database (NED) which is operated by Jet Propulsion Laboratory, California Institute of Technology, under contract with the National Aeronautics and Space Administration. We have also made use of the Lyon-Meudon Extragalactic Database (LEDA), supplied by the LEDA team at the Centre de Recherche Astronomique de Lyon, Observatoire de Lyon. We acknowledge the use of CfA Supernova Archive, which is funded in part by the National Science Foundation through grant AST 0907903; the Online Supernova Spectrum Archive (SUSPECT), initiated and
maintained at the Homer L. Dodge Department of Physics and Astronomy, University of Oklahoma; and  Weizmann Interactive Supernova Data Repository (WISeREP) maintained by the Weizmann Institute of Science computing center.

\label{lastpage}
\end{document}